\documentclass[lettersize,journal]{IEEEtran}
\usepackage{amsmath, amsfonts, amssymb}
\usepackage{multirow}
\usepackage{multicol}
\usepackage{booktabs}
\usepackage{graphicx}
\usepackage{subcaption}
\usepackage{xcolor}
\usepackage{wrapfig}
\usepackage{pifont}
\usepackage{capt-of}
\usepackage{colortbl}
\usepackage[para]{footmisc}
\usepackage{hyperref}
\usepackage[affil-it]{authblk}

\newcommand{\datasetname}{EPIC-SOUNDS}
\newcommand{\cmark}{\ding{51}}%
\newcommand{\xmark}{\ding{55}}%

\title{\datasetname: A Large-Scale Dataset \\of Actions that Sound}
\author{Jaesung Huh$^{1*}$, Jacob Chalk$^{2*}$, Evangelos Kazakos$^{3}$, Dima Damen$^{2}$, Andrew Zisserman$^{1}$}

\affil{$^{1}$Visual Geometry Group, 
Department of Engineering Science, University of Oxford, UK \\
$^{2}$Department of Computer Science, University of Bristol, UK \\
$^{3}$ CIIRC, Czech Technical University in Prague, Czech Republic\\
\normalfont \url{https://epic-kitchens.github.io/epic-sounds/}}

\IEEEpubid{0000--0000/00\$00.00~\copyright~2021 IEEE}
\begin{document}
\twocolumn[{%
  \renewcommand\twocolumn[1][]{#1}%
  \maketitle
  \begin{center}
    \centering
    \captionsetup{type=figure}
    \includegraphics[width=1\linewidth]{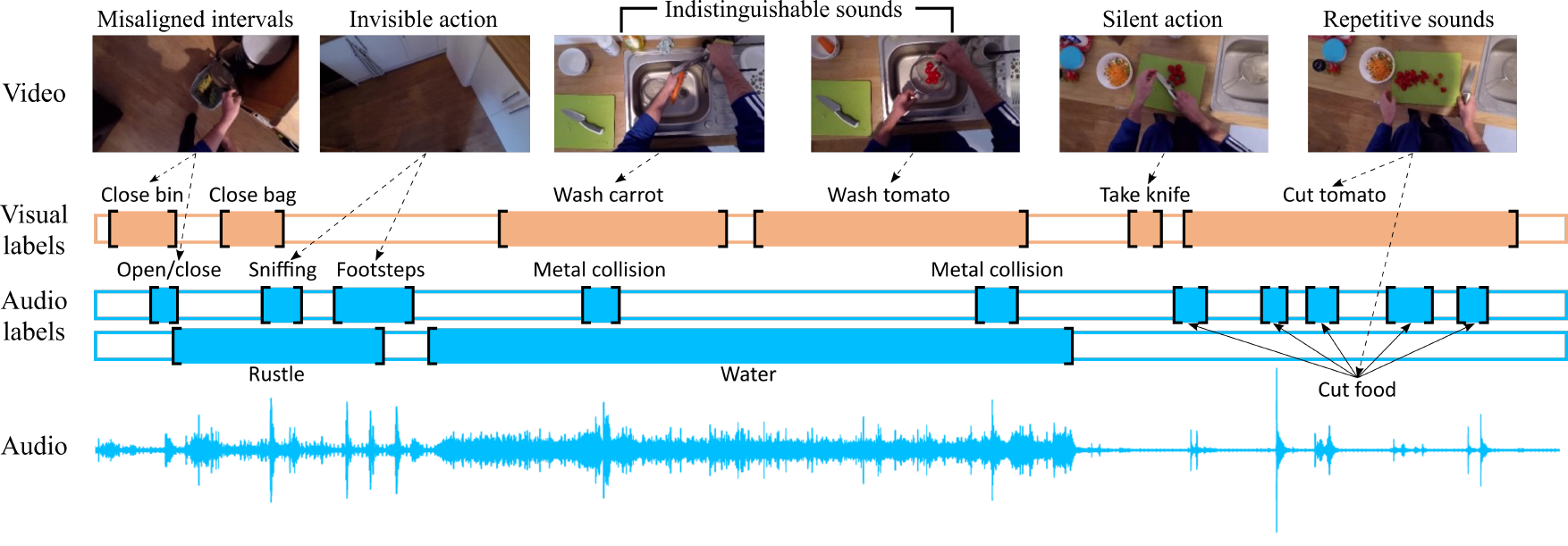}
    \captionof{figure}{Sample video with corresponding audio from EPIC-KITCHENS-100~\cite{Damen2020RESCALING}. We compare the already published  \textbf{visual labels} with our collected \datasetname\ \textbf{audio labels}. We demonstrate the differences between the modality annotations, both in temporal extent and class labels, highlighting: \textbf{Misaligned intervals:}  temporal boundaries are distinct;  \textbf{Invisible action:} action not seen in the video, but which produces distinct sounds (0-to-1 matching); \textbf{Indistinguishable sounds:} sounds from two distinct visual actions, but are audibly inseparable; \textbf{Silent~action:} visual action that does not have audible sounds (1-to-0); and visual actions containing multiple \textbf{repetitive sounds} (1-to-N).}
    \label{fig:teaser}
\end{center}%
}]

\begin{abstract}
We introduce \datasetname, a large-scale dataset of audio annotations capturing temporal extents and class labels within the audio stream of the egocentric videos.
We propose an annotation pipeline where annotators temporally label distinguishable audio segments and describe the action that could have caused this sound.
We identify actions that can be discriminated purely from audio, through grouping these free-form descriptions of audio into classes.
For actions that involve objects colliding, we collect human annotations of the materials of these objects (e.g.\ a glass object being placed on a wooden surface), which we verify from video, discarding ambiguities.
Overall, \datasetname\ includes 78.4k categorised segments of audible events and actions, distributed across 44 classes as well as 39.2k non-categorised segments. 
We train and evaluate state-of-the-art audio recognition  and detection  models on our dataset,  for both audio-only and audio-visual methods. We also conduct analysis on: the temporal overlap between audio events, the temporal and label correlations between audio and visual modalities, the ambiguities in annotating materials from audio-only input,  the importance of audio-only labels and the limitations of current models to understand {\emph{actions that sound}}.
\end{abstract}

\newpage

\makeatletter{\renewcommand*{\@makefnmark}{}
\footnotetext{$^*$Equal technical contribution.}\makeatother}

\section{Introduction}
\label{sec:intro}
Humans perceive objects and actions through multiple senses, especially vision and audition~\cite{smith2005development}. Inspired by this, a plethora of works aim to solve various video understanding tasks, such as action recognition~\cite{gao2020listen,nagrani2020speech2action,Nagrani21c}  and detection~\cite{tian2018audio, bagchi2021hear}, by fusing the two modalities. These attempts are especially common for egocentric video datasets due to the camera's close proximity to the ongoing actions resulting in clearer inputs, both visually and audibly. 
Research has shown improved performance by using audio and video jointly in egocentric data~\cite{kazakos2019TBN, kazakos2021MTCN, ramazanova2022owl, xiong2022m}. 

\IEEEpubidadjcol
In general, these works make two key incorrect assumptions: First, that the visual and auditory events temporally coincide; Second, that a single set of classes can be used 
for both modalities, typically derived from vision. In practice, visual and auditory events exhibit varied levels of both temporal and semantic congruence, thus violating these assumptions  (See Figure~\ref{fig:teaser}). In the case of actions such as `close bin', the onset of the visual event can be defined as the time that the person grasps the handle, whereas the onset of the audio event is delayed to the moment when the lid of the bin slams. 
Some actions are audibly indistinguishable, e.g.\ `wash carrot' vs `wash tomato', as it is impossible to determine which vegetable is being washed through sound alone. 
Consequently, using the visual temporal labels as targets for training an audio classifier is often a flawed endeavour -- the resulting audio classifier will not be able to discriminate all of the visual events; and many audio labels that could provide supervision for training are missed.
Based on these observations, we crowdsource temporal and semantic labels for the audio of EPIC-KITCHENS-100 that are distinct from the visual ones. 

However, as evidence suggests~\cite{vanderveer1979ecological}, humans perform poorly at recognising objects and events using audio alone, making their annotation using only audio challenging. Due to the lack of sufficient information in audio for inferring fine-grained properties of events, humans tend to use vague terms for describing them; \emph{e.g}.\  when the interaction from the collision of two objects is indistinguishable from audio, annotators often describe the associated event as `clang' or `bang'. To alleviate this, we further augment these semantics with the \emph{materials} of the objects that interact. 
We verify these from the video, discarding incorrect audio-only material annotations.

In summary, we introduce \datasetname, a large-scale dataset of daily-life sounds, derived from the audio of EPIC-KITCHENS-100. \datasetname\ contains 78,366 categorised sound events spanning over 44 categories, as well as 39,187 non-categorised sound events, totalling 117,553 sound events across 100 hours of footage collected  in 700 videos  from 45 home kitchens. The sound classes are based on descriptions from only listening to audio, thus suitable for problems in acoustics such as sound recognition and sound event detection.

In this paper, we begin by introducing the related works to \datasetname\ (Section~\ref{sec:related_work}). 
We then introduce \datasetname\ (Section~\ref{sec:epic_sound}) followed by detailing our data collection pipeline (Section~\ref{sec:pipeline}).
We present an extensive analysis of the interplay between audio and visual modalities, as well as the complexity of collecting sounds for material based collisions. (Section~\ref{sec:av-analyis}).
We release two challenges on \datasetname: sound recognition, classifying a sound event given its start and end time, and sound detection, both localising and classifying all sound events within an untrimmed video.  Additionally, we provide strong baseline results for these challenges, using both audio-only and audio-visual approaches (Section~\ref{sec:experiments}).
We conclude our work by reflecting on its contributions and provide an overview of the already achieved impact of \datasetname\ since it's release (Section~\ref{sec:conclusion}).

\begin{table}[t]
\caption{Dataset Comparison. \textbf{A}: Audio. \textbf{V}: Video. \textbf{T}:~Temporal annotations. We showcase that \datasetname, and more recently the Perception Test, are the only datasets with distinct classes for audio and video modalities~(\textbf{D}). We report categorised segments of \datasetname\ here. }
\label{table:existingdata}
\centering
\footnotesize
\resizebox{1\linewidth}{!}{
\begin{tabular}{ l r r r r r r r }
\toprule
\textbf{Name} & \textbf{Source} & \textbf{$\#$hrs} & \textbf{$\#$seg.} & \textbf{$\#$cls} & \textbf{Modality} & \textbf{T} & \textbf{D}\\ 
\midrule
DESED~\cite{Turpault2019_DCASE} & real + synth. & 43h & 8k & 10 & \textbf{A} & \cmark & N/A\\
L3DAS21~\cite{guizzo2021l3das21} & synth. & 15h & 23k & 14 & \textbf{A} & \cmark & N/A\\
URBAN-SED~\cite{salamon2017scaper} & synth. & 30h & 50k & 10 & \textbf{A} & \cmark & N/A\\
TUT 2016~\cite{mesaros2016tut} & real & 2h & 6.3k & 18 & \textbf{A} & \cmark & N/A\\
AudioSet~\cite{audioset} & YouTube & 5833h & 1.8M & 632 & \textbf{A + V} & \xmark & \xmark\\
VGG-Sound~\cite{chen2020vggsound} & YouTube & 550h & 200k & 309 & \textbf{A + V} & \xmark & \xmark\\
SSW60~\cite{ssw602022eccv} & real & 25.7h & 9.2k & 60 & \textbf{A + V} & \xmark & \xmark\\
LLP~\cite{Tian_2020_ECCV} & YouTube & 33h & 19.4k & 25 & \textbf{A + V} & \cmark & \xmark\\
Perception Test~\cite{patraucean2023perception} & real & 68.9h & 113k & 16 & \textbf{A + V} & \cmark & \cmark\\
\midrule 
 \textbf{\datasetname} & home kitchens & 100h & 78.4k & 44 & \textbf{A + V} & \cmark & \cmark\\ 
 \bottomrule
\end{tabular}} 
\normalsize
\end{table}

\section{Related Work}
\label{sec:related_work}
\noindent\textbf{Sound event detection datasets}. Sound Event Detection (SED) is the task of detecting the onset and offset of audio events as well as recognising the event within the detected boundaries. 

SED datasets~\cite{Turpault2019_DCASE, guizzo2021l3das21, salamon2017scaper,mesaros2016tut} are similar to \datasetname\ as these include annotations of temporal boundaries of events, whereas sound recognition datasets~\cite{piczak2015dataset, moreaux2019benchmark, fonseca2022FSD50K} do not. Nevertheless, they differ from \datasetname\ in several aspects. First, they are of smaller scale making the training of modern architectures impractical. Second, \cite{Turpault2019_DCASE, guizzo2021l3das21, salamon2017scaper} contain synthetic audio, and therefore models trained on these datasets generalise poorly to real recordings. Third,~\cite{Turpault2019_DCASE,guizzo2021l3das21,salamon2017scaper,mesaros2016tut} contain sounds associated with generic scenes and events, whereas \datasetname\ focuses on fine-grained sounds generated from diverse audible events in 45 home kitchens.

\noindent\textbf{Audio-visual datasets}. 
We compare \datasetname\ to publicly available sound recognition or detection datasets in Table~\ref{table:existingdata}.
AudioSet~\cite{audioset} is the largest audio-visual dataset of audio events with 2.1M clips and 527 annotated classes, while VGG-Sound~\cite{chen2020vggsound} contains 
over 200K video clips and 300 audio classes. They are both collected from YouTube and each audio clip is 10s long. 
 Both do not have temporal annotations for events, and importantly, a single set of annotations is collected for both modalities. 
 The LLP dataset~\cite{Tian_2020_ECCV} is similar to ours, in that both visual and auditory events are annotated independently, providing separate temporal segments. However, unlike ours, both modalities still share the same label set. Also, LLP is of smaller scale and contains diverse events while \datasetname\ focuses on sounds resulting from actions. 

Closest to ours is the Perception Test~\cite{patraucean2023perception}, which also provides distinct timestamps and labels for both the audio and visual modalities. However, the class diversity is smaller with 16 audio classes versus our 44 audio classes. The videos in the Perception Test are also significantly shorter (23s compared to our average of 514s; 8 minutes and 34 seconds).

\noindent\textbf{Fine-grained audio-visual datasets}. 
The PACS dataset~\cite{yu2022pacs} focuses on understanding the physical common sense attributes of objects shown in the video, similar to our `material' based annotation procedure. However, these attributes are distinguished by 13.4K question-answer pairs; displaying the video with and without audio, and then querying a variety of physical properties. SSW60~\cite{ssw602022eccv} consists of 31K images, 3.8K audio and 5.4K videos of  60 species of birds, proposed to facilitate works on fine-grained categorisation using audio-visual fusion. 
Both datasets do not contain temporal annotations of sounds. 

\section{\datasetname: dataset statistics}
\label{sec:epic_sound}

\noindent \textbf{EPIC-KITCHENS-100.} EPIC-KITCHENS-100~\cite{Damen2020RESCALING} is a large-scale egocentric audio-visual dataset which contains 100 hours of videos containing unscripted daily activities and object interactions in people's kitchens. It consists of  700 videos and  89,977 segments describing visual actions that occur. Actions consist of verb and noun labels, where there are 97 verb classes and 300 noun classes.  The average action length is 2.6s.  Since these actions are based only on video, we emphasise that we do not refer to any of these labels during the annotation process.

\begin{figure*}[t]
    \centering
    \includegraphics[width=\linewidth]{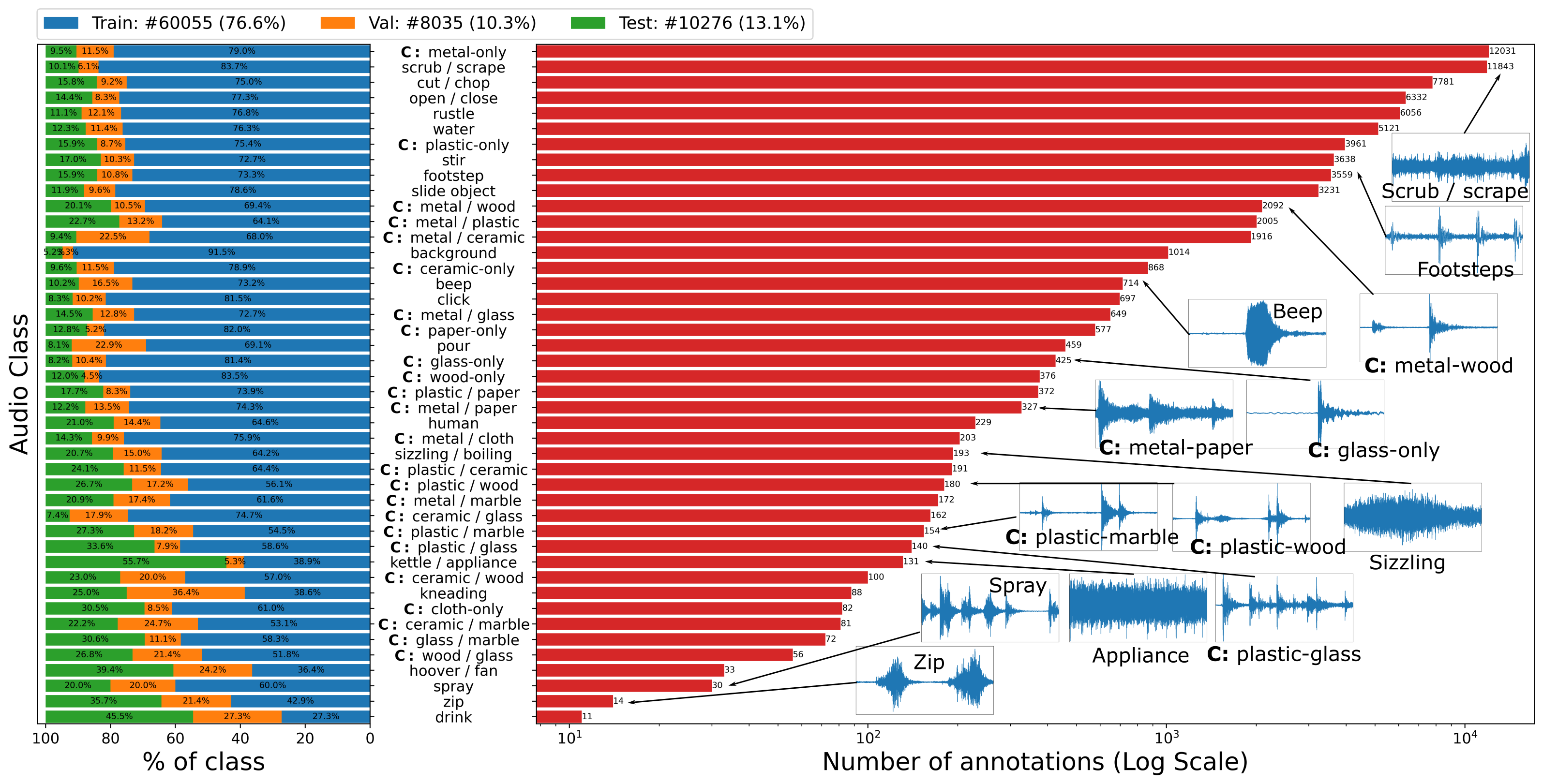}
    \caption{Left: The distribution of each audio class across the \datasetname\ dataset splits. Right: Class frequencies showcasing the long-tail distribution. \textbf{C:} represents a collision-based sound between objects of the same or two distinct material types.}
    \vspace*{-16pt}
    \label{fig:class_dist}
\end{figure*}

\noindent \textbf{\datasetname.} The dataset consists of 78,366 categorised temporal annotations with an average length of 4.9s, distributed across 44 classes. We match the train / validation / test splits from EPIC-KITCHENS-100, giving the per-class proportion across splits in Figure~\ref{fig:class_dist}~(left). 

\begin{figure}[t]
    \centering
    \includegraphics[width=\linewidth]{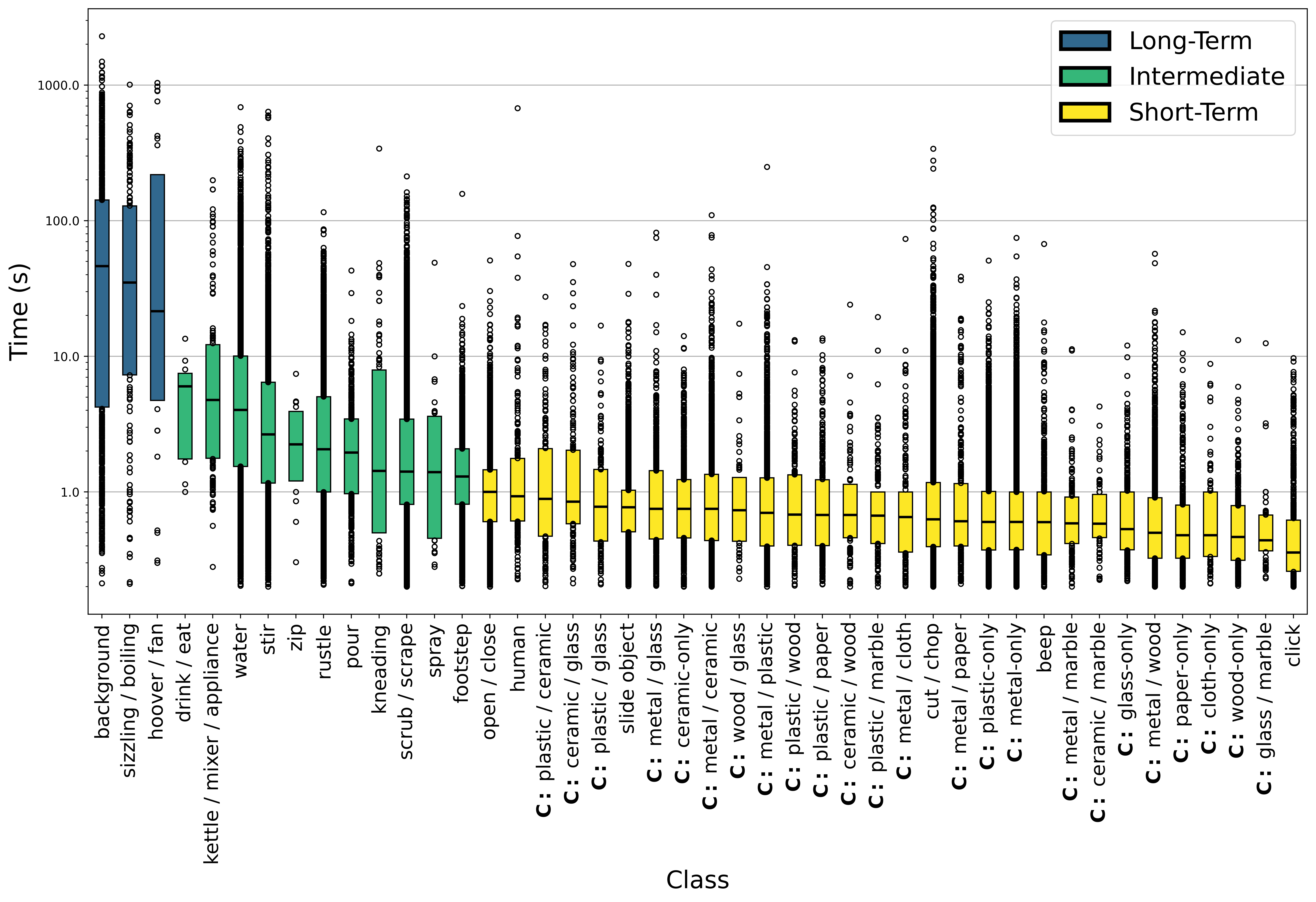}
    \caption{Box plot for the lengths of the annotations over classes, ordered by the median of their lengths. The majority of the classes, 30 (68\%) are short-term, 11 (25\%) are intermediate classes and only 3 (7\%) are considered long-term (median $>$ 10s). \textbf{C:} collision-based sounds between objects of the same or two distinct material types.}
    \label{fig:annotation_length}
\end{figure}

 Class frequency is also shown in Figure~\ref{fig:class_dist}~(right), highlighting that \datasetname\ is naturally long-tailed. We also visualise the waveforms for a sampled subset of the classes. Here, there are both classes which produce waveforms consistent with short-term, percussive sounds such as all the collision-based classes, as well as long-term sounds e.g.\ sizzling.
We also visualise the length of the annotations distributed across the classes in Figure~\ref{fig:annotation_length}. Here, we sort the classes by the median of their lengths, $\Tilde{t}$, and distinguish three categories: long-term ($\Tilde{t}\geq10s$); intermediate ($1s<\Tilde{t}<10s$); and short-term ($\Tilde{t}\leq1s$) classes. Long-term classes relate to lengthier activities, such as cooking and hoovering. In the intermediate classes, there are sounds such as scrub / scrape, or rustle, and then near instantaneous / percussive sounds in the short-term category, including all collision-based classes.

\section{Data collection Pipeline}
\label{sec:pipeline}

The data collection process is conducted through the collection of temporal segments of distinct sounds, described by free-form vocabulary, followed by clustering generic sound categories into distinct classes. This section details this process, as well as post-processing steps taken to refine the results.

\subsection{Data collection of labelled temporal segments}
\label{subsec:temporal}
The objective is to annotate all the distinctive audio events that occur  across all the videos in EPIC-KITCHENS-100.  The annotation consists of the temporal interval of the event, together with a free-form text description.
 As the video length in this dataset varies greatly, from 30 seconds to 1.5 hours,  we trim the videos into a series of manageable lengths for annotations of 3-4 minutes. 
We deem our decision to only provide the audio stream as a key step so the annotators focus on the temporal bounds of the acoustic event alone, rather than being biased by visual and contextual information present in the video stream (consider the `misaligned intervals' example shown in Figure~\ref{fig:teaser}, where visual and auditory temporal segments do not align for the same event). However, the annotators are provided with the plotted audio waveform to act as a visual guide to assist in targeting specific audio signatures and streamline the annotation process. 

\begin{figure}[t]
    \centering
    \includegraphics[width=\linewidth]{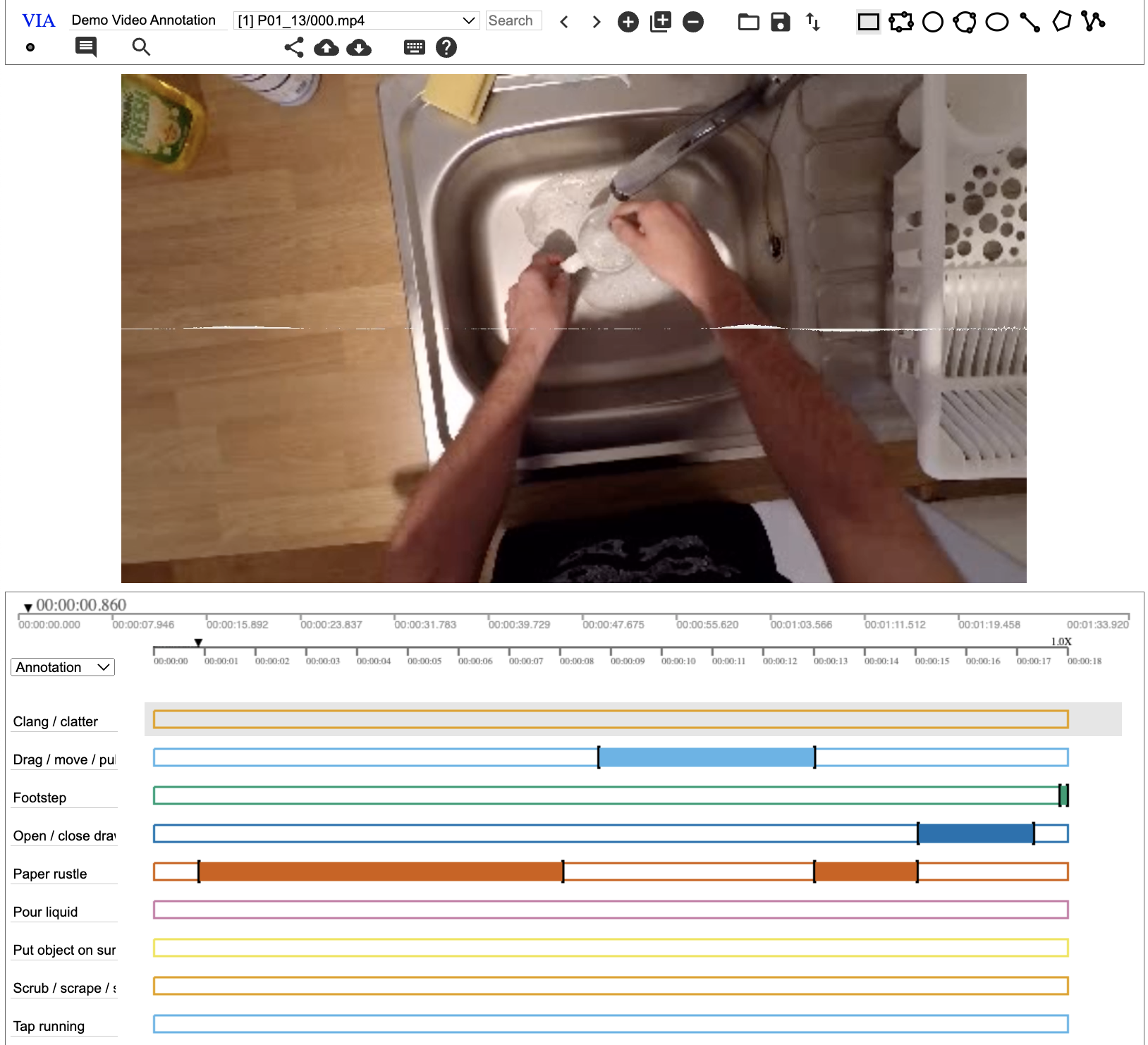}
    \caption{Annotation interface by customising the VIA tool~\cite{dutta2019vgg} to annotate the time interval and semantic label of each distinctive sound the annotator hears. At the top, a single static frame allows understanding the context of the video.}
    \label{fig:via_interface}
\end{figure}

\begin{figure}[t]
    \centering
    \begin{subfigure}[t]{0.49\linewidth}
        \centering
        \includegraphics[height=1.8\linewidth]{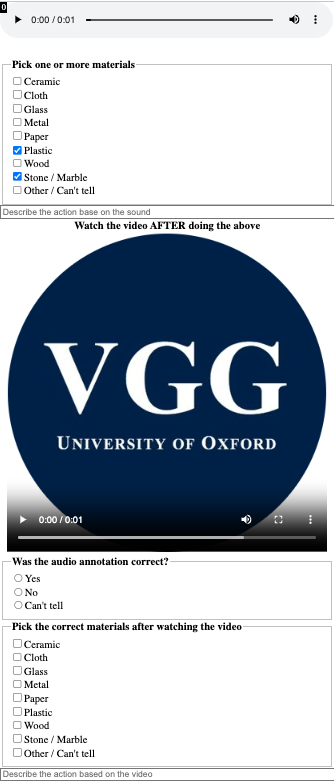}
        \caption{}
        \label{fig:material_interface_hidden}
    \end{subfigure}
    \begin{subfigure}[t]{0.49\linewidth}
        \centering
        \includegraphics[height=1.8\linewidth]{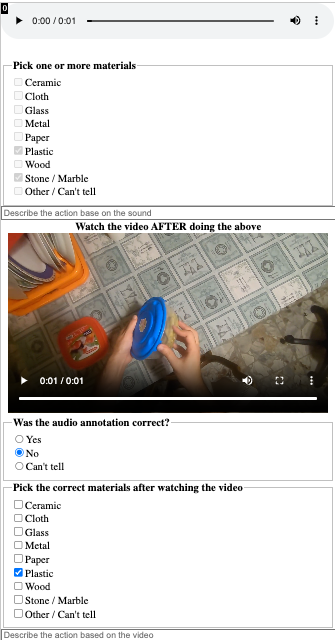}
        \caption{}
        \label{fig:material_interface_visual}
    \end{subfigure}
    \caption{ Customised LISA~\cite{duta20lisa} annotation interface used for annotating the material(s) of the colliding objects based on the given trimmed audio. The interface is in two steps: (a) annotating materials from audio only, and (b) verifying the material from the video input. Annotators cannot change the audio-only prediction after watching the video.}
    \label{fig:lisa_material_interface}
\end{figure}

\noindent \textbf{Annotation process.} We worked with 20 annotators hired from an annotation company.
 For each video segment, we use one annotator to give an initial set of audio annotators and a second annotator to check these.
The first annotator listens to the audio and identifies distinctive audio events, marking their start and end times and assigning a semantic label that reflects the annotator's interpretation of the action or source associated with the sound. 
Annotators use free-form vocabulary, though we provide a reference list of commonly occurring everyday sound labels to guide them which they may optionally select from.
The second annotator performs quality control, e.g.\ reviewing the annotations for any missed events, and makes necessary corrections.

We use a customised version of the VIA tool~\cite{dutta2019vgg} to gather the annotations.
, shown in Figure~\ref{fig:via_interface}. For each unique label description, the VIA tool creates a separate time-line, effectively grouping sequences of the same event. Note that sound events can overlap in time.
If two segments  with the same label  are less than 0.3s apart, we instruct the annotators to merge the two segments as we deem them to belong to the same event. 
Additionally, annotators are asked to identify consistent background sounds (or noise) that occur throughout a large portion of the audio (e.g.\ radio, fan or washing machine). 
 The annotators were asked to tag these as `background'.  
The procedure described thus far resulted in the annotation of 556 distinct sound descriptions.

Humans tend to use abstract words to describe sounds, such as `clang' or `clatter', especially for those generated from the collisions between objects. 

However, humans are able to further understand the materials of these objects that have collided -- for example they can distinguish between two glasses colliding versus a plastic container colliding with a wooden surface.
We thus opt to annotate the material(s) involved in these collision sounds.
We use a customised LISA~\cite{duta20lisa} annotation interface for annotating the material of the objects that collide based on audio (Figure~\ref{fig:lisa_material_interface}).

We instruct annotators to select from a pre-specified list of which {\emph{materials}} are involved in the collision, provided in Table~\ref{tab:materials}. 
These cover all the materials popular in kitchens.
Annotators are encouraged to select one or more materials, or mark the material as indistinguishable by choosing the `Can't tell' option.  
We drop the instances in the latter case -- as we believe these are unhelpful for sound or event understanding tasks. 
However,  some material sounds might be deceiving. 
For example, one might perceive the material collision to be between a glass and a wooden object, but in fact it is food poured into a ceramic container.
We thus ask annotators to then visually verify their material annotations using the corresponding video.
Importantly, annotators have to listen and choose the perceived material first 
(Figure~\ref{fig:material_interface_hidden}),
 
and cannot change these after watching the video 
(Figure~\ref{fig:material_interface_visual}).

Instead, they select the actual materials involved when viewing the video. 
We only retain visually-verified collision sounds -- i.e.\ materials correctly perceived from the audio only, then verified from the visual observation. 
We choose all collision material labels for which at least 40 examples are present.
As a result, abstract labels related to collision (e.g.\ `clang/clatter’, `put objects on surface’) are clustered into 24 sound categories describing the materials involved, such as \textbf{C}: metal-only, or \textbf{C}: plastic-wood,  where \textbf{C} indicates collision-based classes.

\begin{table}[t]
\caption{Material options for collision sounds.  We note \textbf{\# of time} each material was \textbf{selected} in collision sounds, and discard the sounds annotated with `Others' or `Can't tell'.}
\label{tab:materials}
    \footnotesize
    \centering
    \resizebox{1\linewidth}{!}{
    \begin{tabular}{llc}
    \toprule
    \textbf{Material}       & \textbf{Example objects}        & \textbf{\# of times selected}\\
    \midrule
    Metal          & metal or stainless steel&        15523 \\
    Plastic        & plastic bowl, plastic container&  5464 \\
    Ceramic        & ceramic cup, plate      &   2634     \\
    Wood           & wooden spatula, wooden table&      2408\\
    Paper          & kitchen roll, cardboard boxes  & 1253 \\
    Glass          & wine glasses, glass cup&   1248      \\
    Stone / Marble & kitchen worktops, marble tables &    377  \\
    Cloth          & towels, teatowels, clothes&     257\\
    Others         & materials not listed above (e.g.\ food)&  3596 \\
    Can't tell     & cannot determine the material&  10030 \\
    \bottomrule  
    \end{tabular}}
\end{table}

\subsection{Post-processing Annotations}
\label{subsec:post-processing}

\noindent \textbf{From labels to classes.} We post-process the audio labels to fix spelling errors and group semantic equivalences. 
For example, sounds like `buzzer', `beep' and `alarm' are grouped into one {\emph{beep}} class.
Similarly, sounds described by the verbs `wipe', `scour', `scrape' and `scrub' are also grouped into a single class.
We also manually review tail instances to determine whether these form novel classes or should be merged with others.  In cases where the description was not meaningful, the categorised annotation is dropped.  For example, the sound `spray' was considered a meaningful tail instance of an action that sounds. In contrast, the label `dog barking' was discarded as it is not relevant to our context. This produces the 44 audio classes, as shown in Figure~\ref{fig:class_dist}.

\noindent \textbf{Error checking audio classes.} Due to differences in sound perception between annotators, some errors exist amongst the classes. For example, where one annotator hears a drawer being pulled and hence labels `open / close', another may hear `drag object' for a similar audio. To resolve such errors, we manually review each of the labels in the test and validation set.

\begin{figure}[t]
    \centering
    \begin{subfigure}[t]{0.32\linewidth}
        \centering
        \includegraphics[height=1.8\linewidth]{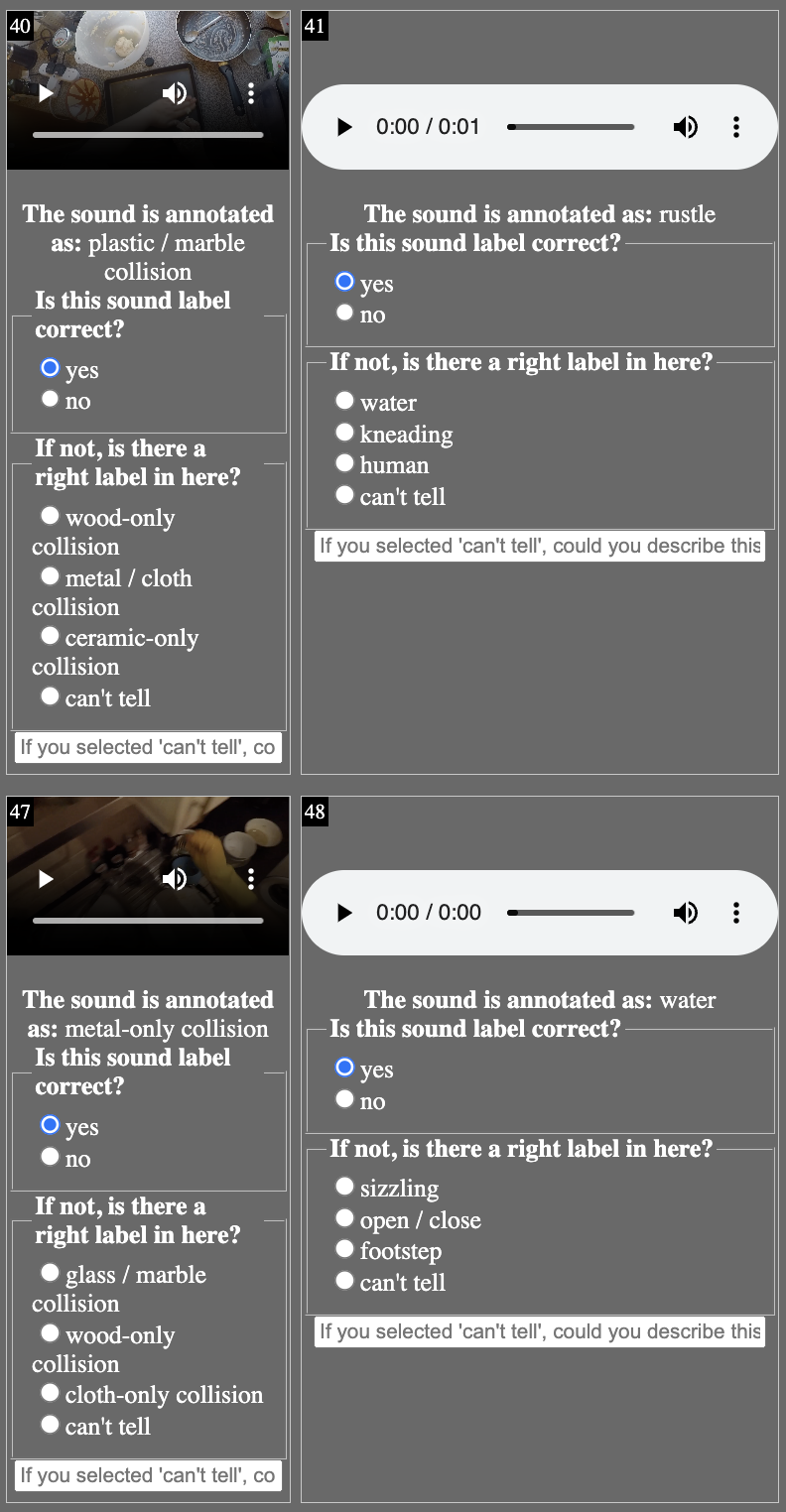}
        \caption{}
        \label{fig:manual_checking_interface}
    \end{subfigure}
    \begin{subfigure}[t]{0.32\linewidth}
        \centering
        \includegraphics[height=1.8\linewidth]{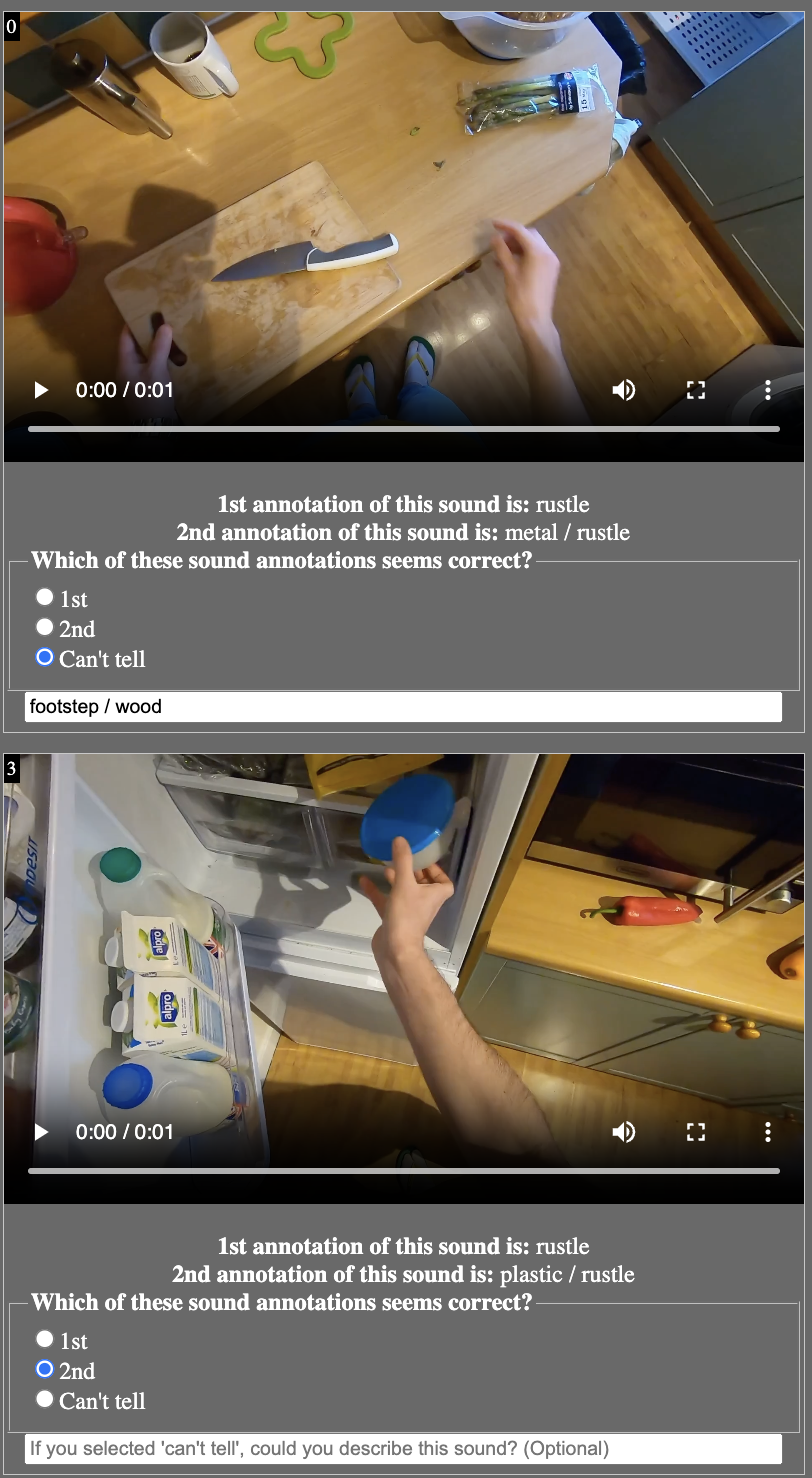}
        \caption{}
        \label{fig:disagreement_interface}
    \end{subfigure}
    \begin{subfigure}[t]{0.32\linewidth}
        \centering
        \includegraphics[height=1.8\linewidth]{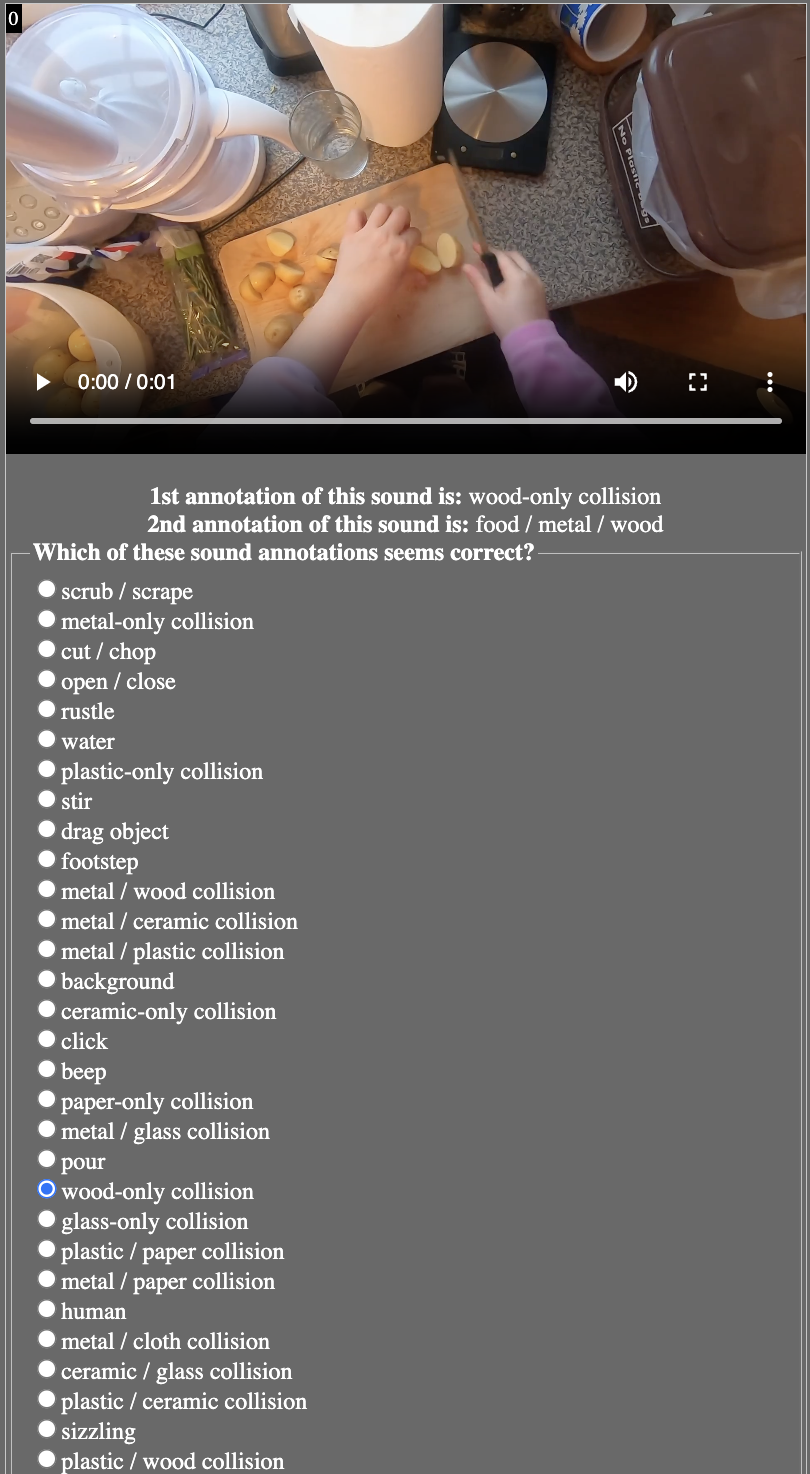}
        \caption{}
        \label{fig:non_trivial_interface}
    \end{subfigure}
    \caption{ Customised LISA~\cite{duta20lisa} annotation interfaces used for: (a) manually checking the trimmed event labels, (b) correcting disagreeing samples from the manual checking stage and (c) choosing between different annotations.
    }
    \label{fig:lisa_correction_interface}
\end{figure}

The procedure is as follows: first, using the interface shown in Figure~\ref{fig:manual_checking_interface}, we ask the annotators to manually review all the validation / test samples. 
For non-collision sounds, we provide only the audio, again to avoid any visual bias during annotations, and ask annotators to verify the audio labels.
We further test this in a closed-form QA setting, again asking an annotator to select from 4 sound labels, out of which one is the previously-labelled sound.
If the label is correctly selected again, it will be considered the final audio ground-truth.
For collision-based sounds, it is difficult to ascertain the material of objects involved in collision sounds from audio alone, we instead provide the annotators the raw audio-visual footage to perform their correction.
For both types of sounds, we also provide a free-form text-box to allow annotators to provide their own descriptions, should they believe these are better for describing the sound.
Additionally, we include a `can't tell' option if the true sound is difficult to describe.

Following the initial round of corrections, we focus on the instances where the annotators adjust the initial labels.
Using the interface shown in Figure~\ref{fig:disagreement_interface}, we ask a different set of annotators to choose between these two annotations for the correct one, also providing them with a `can't tell' or `neither' option, where they can provide an additional third annotation.
All label corrections are manually reviewed before adopted. 
We use an interface particularly to check non-trivial conflicting labels, shown in Figure~\ref{fig:non_trivial_interface}.

For the training set, we utilise the overlaps between audio segments and visual segments to select the samples for reviewing. 
We deem the use of the visual labels acceptable for error correction, as the annotation process is complete. Thus, utilising the visual labels for post-processing no longer compromises the issues stated in Figure~\ref{fig:teaser}. 
We review all audio classes for which there exists a {\emph{mapping}} to visual classes in EPIC-KITCHENS-100. We identify two types of mapping, trivial; the audio class itself already exists as a visual class e.g.\ `scrub', and relational; the audio class does not exist as a visual class itself but can be semantically mapped to one or more of the visual classes, such as the audio class `click' relating to the verb `turn on / off' or the noun `light switch'. 

We manually review all cases where an audio annotation overlaps with a \emph{different} visual action -- e.g.\ a `scrub' audio class with one overlapping visual annotation of the class `open / close'. 
We use this additional filtering to correct the audio classes.

We run this error-checking cycle multiple times to ensure that all incorrectly classified instances are accounted for.

\noindent \textbf{Non-categorised audio events.}
As a result of post-processing, there are audio events that we recognise a sound exists but no semantic label could be given. These are samples we either could not assign class labels through the various correction stages, or collision sounds for which they could not be visually verified. We release the free-form descriptions and temporal boundaries of these 39,187 samples as \textit{non-categorised}.

\begin{figure}[t]
    \centering
    \includegraphics[width=\linewidth]{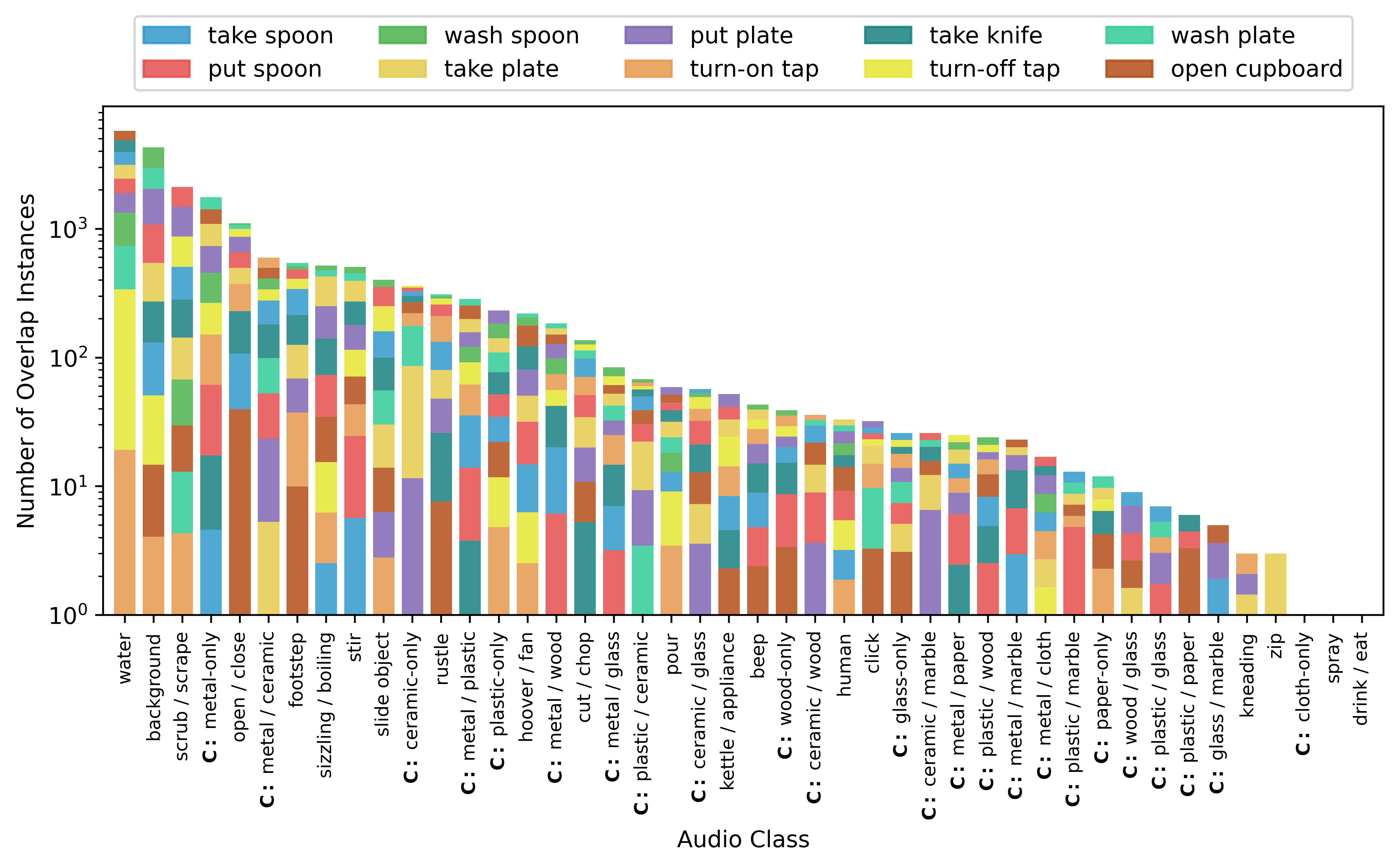}
    \caption{
     Bar chart showing the diversity of temporal overlaps between visual and audio classes. The proportion of overlapping instances for each of the top-10 most frequent temporally overlapping visual classes divides the bar of each audio-class. Note, the plot is log-scaled, but bar division is linearly scaled.
    }
    \label{fig:visual_proportions}
\end{figure}

\begin{figure}[t]
\centering
\begin{subfigure}[t]{\linewidth}
    \includegraphics[width=\linewidth]{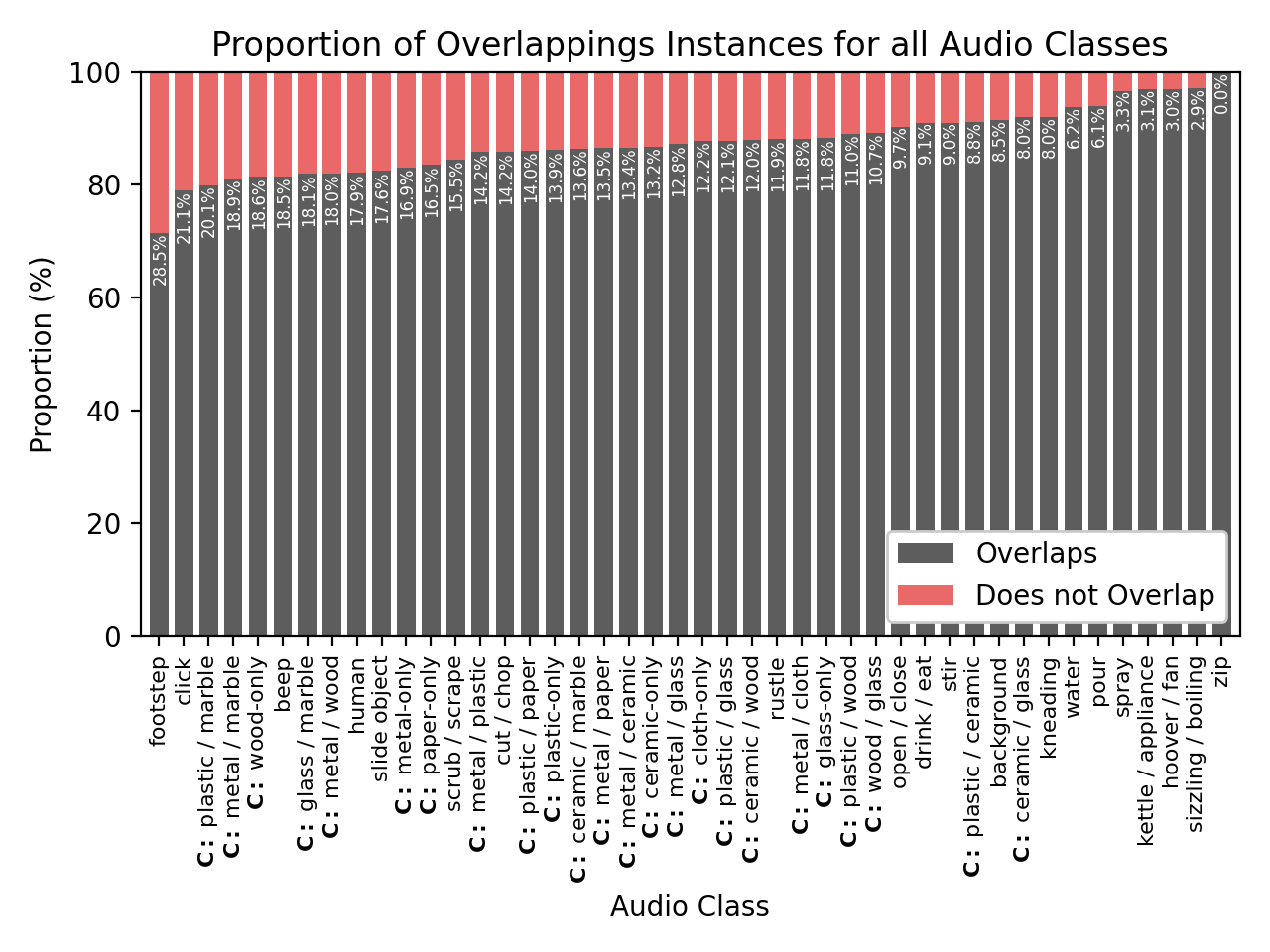}
\end{subfigure}
\hfill
\begin{subfigure}[t]{\linewidth}
   \includegraphics[width=\linewidth]{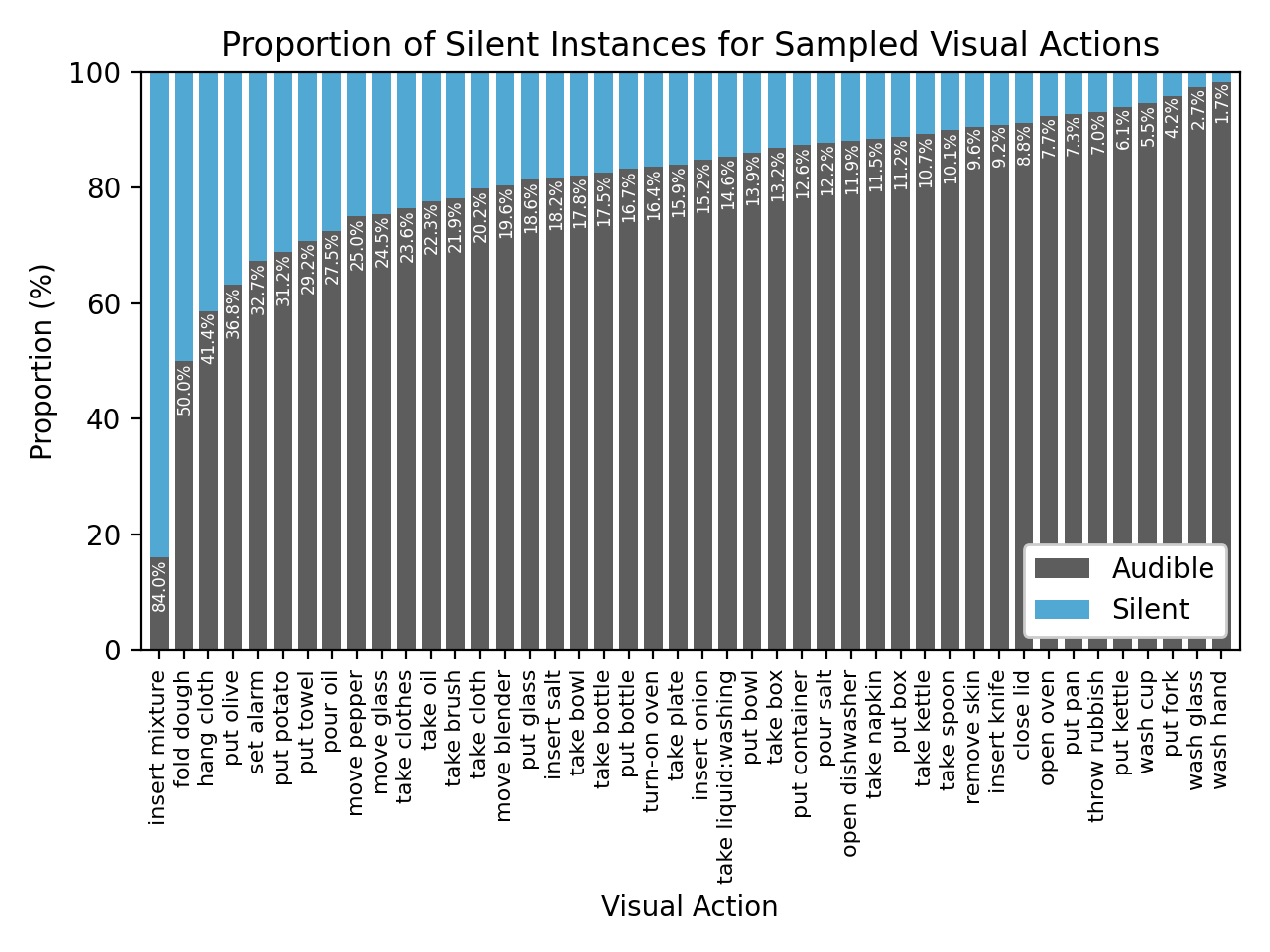}
\end{subfigure}
\caption{
     Bar charts showing the proportion of audio classes that have no temporal overlap with visual classes i.e.\ no clear visual signal (top), and visual classes that have no temporal overlap with audio classes i.e.\ no audible signal and are thus considered silent visual actions (bottom).
}
\label{fig:label_analysis}
\end{figure}

\section{Audio-Visual Analysis}
\label{sec:av-analyis}
Having collected audio-only annotations, forming \datasetname, we now compare these to the already collected visual annotations for EPIC-KITCHENS~\cite{Damen2020RESCALING}. These are start-end times and action labels for the visual stream. We refer to these accordingly as visual events, and refer to the \datasetname\ annotations as auditory events.
In this section, we perform an extensive analysis, investigating the interplay between the audio and visual events and how knowledge of one can benefit the knowledge of the other modality.

\subsection{Overlap of Visual and Auditory Events} 
\label{sec:av-overlap}
When comparing audio to video labels, we reflect on our motivation in Figure~\ref{fig:teaser}. 
For each audio class, we study the  temporal  overlap with visual classes  and make deductions based on such overlaps.
  Figure~\ref{fig:visual_proportions} visualises for each audio class the  number of instances for  the top-10 most frequent  temporally  overlapping visual classes.  Individual bars are divided by the proportion of instances for each overlapping visual class. 
The figure shows the diversity of 
 overlaps across the audio classes and implies that there are  strong associations between  visual and audio classes. For example, the visual class  `open cupboard'  has frequent temporal overlap with  audio classes `open / close' or `footstep', noting that, in many cases, participants walk towards cupboards they are about to open.

Figure~\ref{fig:label_analysis} (top) shows the proportion of audio events  that have no temporal overlap with any  visual event. We hypothesise that no  temporal  overlap implies that the auditory event corresponds to an out-of-view,  or trivial  visual event. 
Conversely,  a temporal overlap between audio and visual classes  typically represents an in-view audio-visual event. 
We see that the top three audio classes 
 with the highest proportion of instances containing no temporal overlap with visual classes are: 
footstep (28.5\%), click (21.1\%) and plastic-marble collision (20.1\%). In these instances, the classes relate to sounds produced by visual actions that  typically  happen off-screen, or are occasionally deemed trivial, such as placing an object or turning on the hob, resulting in missed visual annotations while still producing distinctive auditory signals. 
On the other hand, the three classes which most frequently  have temporal  overlap with visual classes are zip (100\%), sizzling / boiling (97.1\%) and hoover / fan (97.0\%). For zip, these relate to clear visual actions that are frequently annotated (opening / closing a bag), whereas sizzling / boiling and hoover / fan relate to long-form audio which occurs during multiple visual activities, such as pan frying food whilst completing other steps of a recipe, or when an extractor fan is switched on while cooking.

 We also visualise the proportion of visual actions with no temporally overlapping audio event in Figure~\ref{fig:label_analysis} (bottom). 
We hypothesise that no temporal overlap with audio classes corresponds to silent visual actions, whilst the presence of a temporal overlap indicates an audible visual action. 
Here we plot a representative sample of visual actions, across various proportions of silent instances.
We see the top-3 visual classes  with the highest proportion of no temporal overlap with audio classes  (0-to-1) are: insert mixture (84.0\%), fold dough (50.0\%) and hang cloth (41.4\%). Indeed these are actions that seldom produce a discernible sound. Conversely, we see that the top-3 visual actions with  frequent temporal overlap with audio classes  are wash spoon (98.2\%), wash glass (97.0\%) and pour water (95.3\%) where clearly the water will produce clear audio signals that are easily recognisable.

We investigate repeated sounds across a single visual action. We find the top-3 \mbox{many-to-1} audio-to-video classes containing repeated audio sounds (on average) are: cut / chop (2.28-to-1), beep (1.47-to-1), metal / wood collision (1.24-to-1), these relate to actions that have a `stop-start' pattern e.g.\ pauses between chops, button presses on an appliance, or between repetitively moving items in a cutlery drawer or sink.

\begin{figure}[t]
    \centering
    \includegraphics[width=\linewidth]{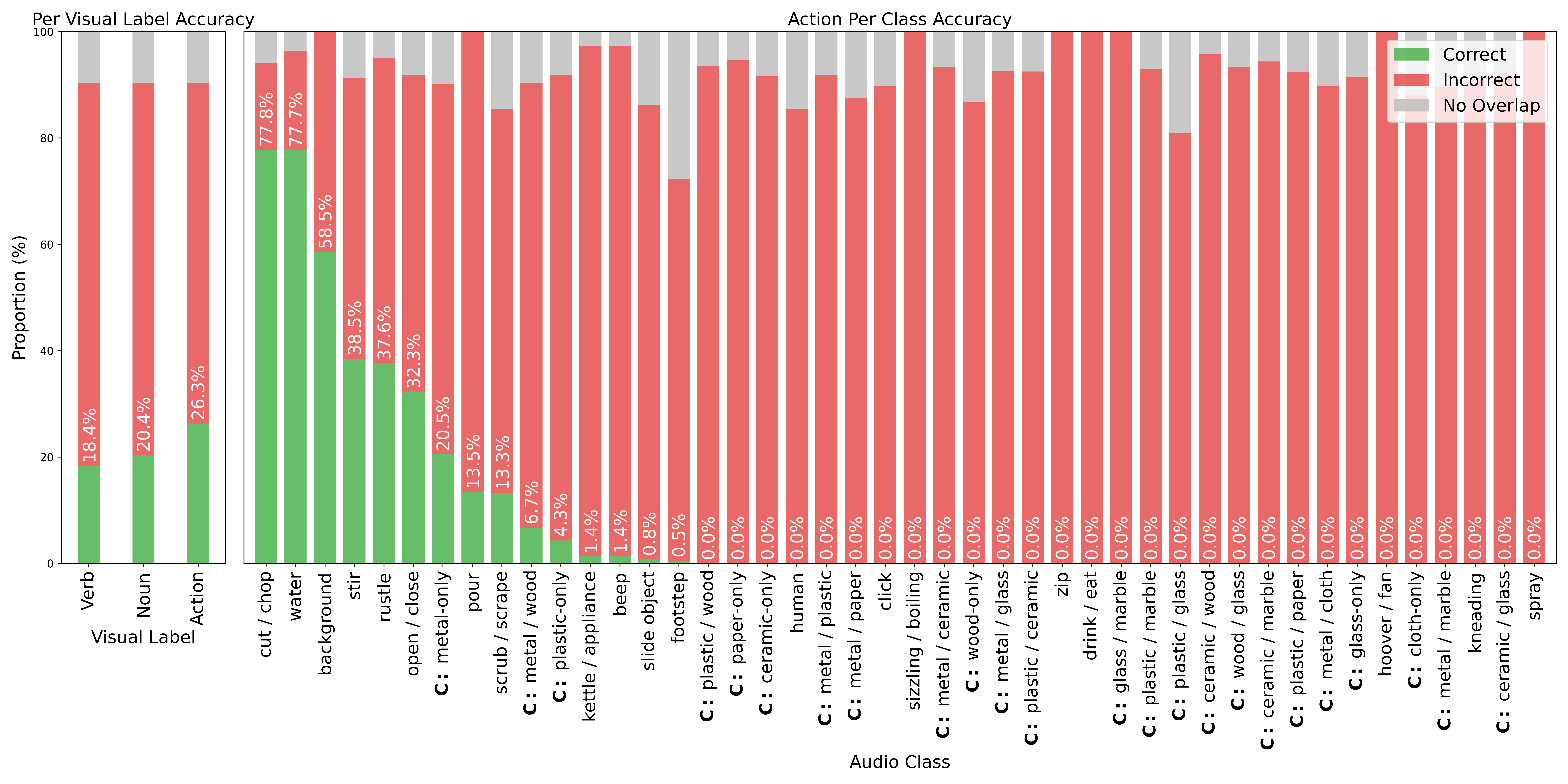}
    \caption{ Bar charts visualising the accuracy on the validation set when using audio-visual overlapping priors. We report visual verb, noun and action accuracy (left) as well as the per-class accuracy for all audio classes using the action priors (right).}
    \label{fig:prior_analysis}
\end{figure}

\subsection{Audio-Visual Prior Analysis} 
\label{sec:prior_analysis}
For this section, we investigate the correlation between the visual and auditory labels. 
Even though visual and audio events differ in their start-end times and labels, some correlations are deterministic in nature. For example, if the visual event is `wash plate', one can guess that there's an auditory sound of `water' in the audio modality.
We study these correlations using a prior analysis.
We compute priors from the training set and see their effectiveness for predicting audio classes in the validation set. 
Specifically, for each visual class in the training set, we find all the overlapping audio instances across the training set.
This produces the prior probability of predicting an audio class given the visual label.
For example, for the visual class `wash plate', in the training set, $p(audio = `water' | visual = `wash\; plate') = 0.8$.

Once the priors are calculated, we use this to calculate the accuracy for the validation set as follows. For a visual event, we consider the overlapping audio instance and its ground truth class, then assign the visual event the visual class with the maximum probability given the prediction of the audio class, as calculated in the training set.
We compare this to the ground truth visual class, evaluating whether the prediction is correct / incorrect or whether no overlapping audio event exists.
Figure~\ref{fig:prior_analysis} (left) reports the overall accuracy for verb, noun and action prediction.
Analogously, we calculate the accuracy for the audio classes as follows.
For each audio event, we consider the overlapping visual action and its ground truth class.
We assign the audio event the class with the highest probability from the prior analysis.
Figure~\ref{fig:prior_analysis} (right) presents these results.
We see that knowing the audio does not produce high accruacy in the visual domain -- accuracy of verb is 18\%, noun is 20\%. Using priors computed from the combination of verb and noun (action) slightly improves performance (26.2\%). 
However, clearly the modalities are quite independent and cannot be predicted from prior analysis
When looking at the audio class accuracy from the visual actions priors, we see that the top-3 correctly classified audio classes are water (62.0\%), open/close (50.6\%) and rustle (50.2\%). These are classes which have clear visual signals, such as the sink when washing, cupboards and drawers and paper/plastic bags.
For 31 out of 44 classes, the accuracy of predicting the audio class from the visual class is 0.

The prior analysis shows that while some labels correlate, it is not possible to predict one modality from the other.

\subsection{Material Analysis} 
\label{sec:material_analysis}

In this section, we investigate human perception of material sounds using our collision sound annotations. 
We limit our scope to sounds produced by eight specific materials listed in Table~\ref{tab:materials}, excluding those labelled as 
`Others' or `Can't tell'.

\begin{figure}[t]
    \centering
    \vspace{-10pt}
    \includegraphics[width=\linewidth]{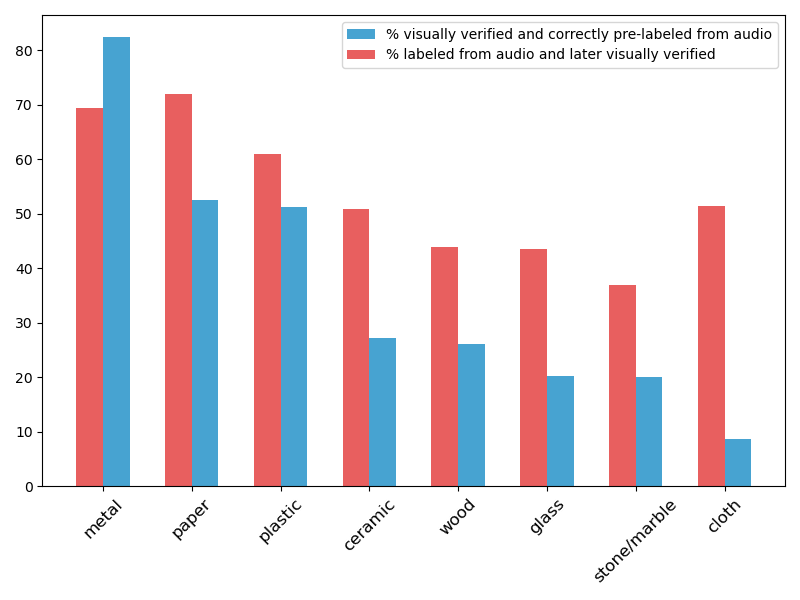}
    \vspace{-10pt}
    \caption{ The recognition accuracy of materials in collision sounds. Red bars refer to the ratio of material recognised purely from audio to materials verified from vision. Blue bars refer to the ratio of visually verified material sounds that were correctly pre-recognised from audio. }
    \label{fig:material_analysis}
\end{figure}

We compute two metrics per material and report the results in
Figure~\ref{fig:material_analysis}. 
Red bars represent the ratio of materials recognised purely from audio to materials verified from vision (\textit{Of those visually verified as material X, how many were pre-labelled as material X?}).
Blue bars represent the ratio of material recognised purely from audio then verified by vision (\textit{Of those labelled as material X only with audio, how many were visually verified as material X?}).
Results show that metal performs highly on both metrics due to its distinctive, resonant sound when struck, which is easily distinguishable from other materials.
Cloth exhibits low performance, especially on the ratio of visually verified cloth sound to the cloth sound recognised only with audio. 
Cloth typically produces muffled, soft sounds upon impact, which are less characteristic and more easily confused with other materials or ambient noise.

Furthermore, our analysis reveals that in \textbf{48.8\%} of instances, annotators responded `yes' to the question ``Was the sound annotation correct?''.
This indicates that for nearly half of the samples, annotators accurately identified all materials generating the collision sounds based solely on audio cues.

\begin{figure}[t]
    \centering
    \vspace{-10pt}
    \includegraphics[width=\linewidth]{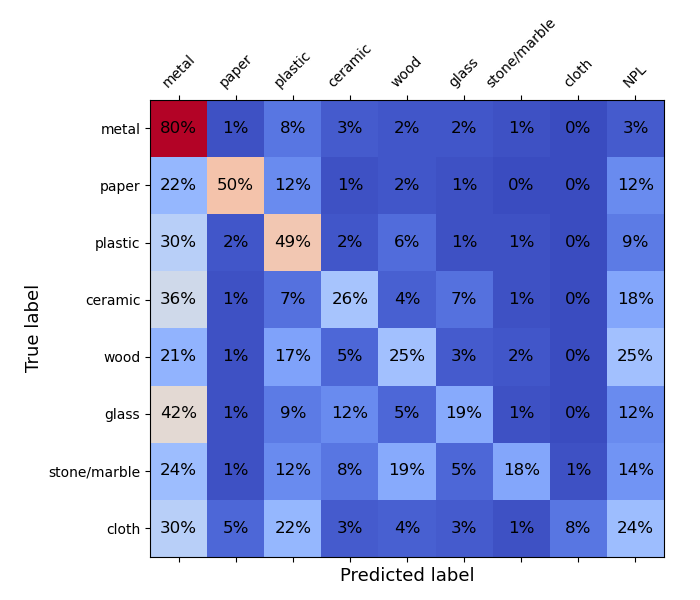}
    \vspace{-10pt}
    \caption{ Material recognition confusion matrix. We adopted Multi-label confusion matrix proposed by \cite{heydarian2022mlcm}. \textbf{NPL} (No Predicted Label) indicates instances where the model failed to predict a material label that is present in the ground truth.}
    \label{fig:material_confusion_matrix}
\end{figure}

 Figure~\ref{fig:material_confusion_matrix} shows the confusion matrix for material recognition from human annotators. Since annotators are capable of labelling multiple materials, we adopt the Multi-Label Confusion Matrix (MLCM) proposed by \cite{heydarian2022mlcm}. \textbf{NPL} (No Predicted Label) denotes cases where a material present in the ground truth was not predicted by the annotator. 
For instance, if the true label is metal / plastic and the annotator only predicts metal, the NPL column for plastic is incremented.
Annotators most accurately identify metal (80\%), paper (50\%), and plastic (49\%). In contrast, stone/marble (18\%) and cloth (8\%) are more difficult to recognise. 
Notably, glass is frequently misclassified as metal (42\%) and other hard materials such as ceramic and plastic are often confused with metal as well.

\section{Challenges and Baseline Results}
\label{sec:experiments}
In this section, we define and experiment on two challenges: sound recognition and sound detection. For each challenge, we evaluate both audio-only models and audio-visual models to measure the complementary nature of the visual modality.

\subsection{Challenge Definitions}

\noindent \textbf{Sound Recognition.} Given an audio segment $S^{i}=[t^{s}_{i},t^{e}_{i}]$, we aim to classify the ongoing sound event within the segment $c^{s}_{i}\in C$, where $C$ is the 44 classes in \datasetname. Note that for this challenge, the start and end time of all segments are known, meaning that all start and end times are given during inference, and the model is only expected to classify the sound. To assess this challenge, we report: top-1 and top-5 accuracy, mean average precision (mAP), mean area under ROC curve (mAUC) and mean per class accuracy (mCA).

\noindent \textbf{Sound Detection.} 
We consider the full untrimmed video $X$, and our aim is to predict all sound event instances in $X$ i.e.\ $\mathbf{\hat{S}}=\{\hat{S}_{i}\}^{N}_{i=1}$ where $\hat{S}_{i}=(\hat{t}^{s}_{i},\hat{t}^{e}_{i}, \hat{c}^{s}_{i})$ specifies a sound detection tuple containing the start and end time of the sound event $(\hat{t}^{s}_{i},\hat{t}^{e}_{i})$ as well as the predicted sound event class $\hat{c}^{s}_{i}$. 
During training, models have access to the ground-truth annotations within $X$ but, unlike in recognition, they no longer have access to the timestamps during inference. 
When evaluating this challenge, we use the mean Average Precision (mAP) metric, which is computed from the mean of the AP values across different IoU thresholds across all classes. If a predicted segment matches a ground truth segment with an Intersection over Union (IoU) greater than the given threshold, it is considered a valid detection. For computing mAP, we average the AP across thresholds $[0.1, 0.2, 0.3, 0.4, 0.5]$.

\noindent \textbf{Task Specific Test Splits.} We divide the \datasetname\ test set into two task-specific sub-sets: i) The Recognition Test split, where we release the timestamps for all actions, but not their labels and ii) The Detection Test split, where we release neither the timestamps or labels, instead only releasing the video IDs where the sounds are present. We select these sub-sets to be roughly equal and non-overlapping in videos. More specifically, the recognition test set contains 5131 sounds, across 44 videos from 11 participants, whereas the detection test set contains 5145 sounds across 23 videos from 9 participants.

\begin{table}[t]
    \footnotesize
    \centering
    \caption{Results of the Baseline Models on the \datasetname\ validation,  recognition test and entire test splits. M: Modality;  L: Linear-Probe; F:~Fine-Tuning.  $*$ uses additional information. (e.g.\ start and end time of neighbouring actions)}
    \label{tab:quantresults}
    \resizebox{1\linewidth}{!}{
    \begin{tabular}{ccccccccc}
        \toprule
         Split & Model & M &  & Top-1 & Top-5 & mCA & mAP & mAUC \\
         \midrule
         \multirow{7}{*}{\rotatebox{90}{\textbf{Validation}}} & Chance & - & -  & 7.71 & 30.95 & 2.29 & 0.023 & 0.500 \\
         & SSAST~\cite{SSAST} & A & L & 28.74 & 64.87 & 7.14 & 0.079 & 0.755 \\
         & ASF~\cite{kazakos_2021_ICASSP} & A &  L & 45.53 & 79.33 & 13.48 & 0.172 & 0.789 \\
         & SSAST~\cite{SSAST} & A &  F & 53.47 & 84.56 & 20.22 & 0.235 & 0.879 \\
         & ASF~\cite{kazakos_2021_ICASSP} & A &  F & 53.75 & 84.54 & 20.11 & 0.254 & 0.873 \\
         & TIM~\cite{Chalk_2024_CVPR} & A+V &  F & \textbf{58.49} & 86.53 & 26.05 & 0.305 & 0.883 \\
         & MTCN$*$~\cite{kazakos2021MTCN} & A+V &  F & 57.50 & \textbf{86.82} & \textbf{26.44} & \textbf{0.314} & \textbf{0.920} \\
         \midrule
         \multirow{7}{*}{\rotatebox{90}{\parbox{1.5cm}{\centering\textbf{Recognition\\ Test}}}} & Chance & - & -  & 7.85 & 31.91 & 2.39  & 0.024 & 0.500 \\
         & SSAST~\cite{SSAST} & A &  L & 29.93 & 66.60 & 7.17 & 0.082 & 0.725\\
         & ASF~\cite{kazakos_2021_ICASSP} & A & L & 45.00 & 78.98  & 15.00  & 0.183 & 0.788 \\
         & SSAST~\cite{SSAST} & A &  F & 53.71 & 84.54 & 22.28 & 0.223 & 0.820 \\
         & ASF~\cite{kazakos_2021_ICASSP} & A &  F &  54.45 & 85.17 & 20.41 &  0.254 & 0.852\\
         & TIM~\cite{Chalk_2024_CVPR} & A+V & F & 55.31 & 85.09 & 24.22 & 0.290 & 0.861 \\
         & MTCN$*$~\cite{kazakos2021MTCN}& A+V & F & \textbf{57.55} & \textbf{87.51} & \textbf{27.09} & \textbf{0.308} & \textbf{0.900} \\
         \midrule
         \multirow{7}{*}{\rotatebox{90}{\parbox{1.5cm}{\centering\textbf{Entire\\ Test}}}} & Chance & -  & - & 7.22 & 30.11 & 2.27  & 0.023 & 0.500 \\
         & SSAST~\cite{SSAST} & A & L & 27.50 & 65.55 & 6.68 & 0.080 & 0.741\\
         & ASF~\cite{kazakos_2021_ICASSP} & A &  L & 44.55 & 78.44  & 14.49  & 0.145 & 0.772 \\
         & SSAST~\cite{SSAST} & A &  F & 53.75 & 83.76 & 20.76 & 0.237 & 0.860 \\
         & ASF~\cite{kazakos_2021_ICASSP} & A &  F &  54.86 & 84.26 & 20.30 &  0.232 & 0.823\\
         & TIM~\cite{Chalk_2024_CVPR} & A+V &  F & 55.53 & 85.35 & 23.72 & \textbf{0.319} & 0.882 \\
         & MTCN$*$~\cite{kazakos2021MTCN} & A+V & F & \textbf{57.96} & \textbf{87.55} & \textbf{26.52} & 0.308 & \textbf{0.908} \\
         \bottomrule
    \end{tabular}}
\end{table}

\subsection{Audio-only Sound Recognition}
Here, we describe how audio-only state-of-the-art sound recognition models perform on classifying \datasetname.

\noindent \textbf{Baselines.} We train and evaluate the Auditory SlowFast (ASF)~\cite{kazakos_2021_ICASSP} and Self-Supervised Audio Spectrogram Transformer (SSAST)~\cite{SSAST} audio encoder networks, with both a linear probe,  i.e.\ by freezing the model weights and only training the classification layer,  and by fine-tuning.  We also compare to a chance baseline.  ASF is pretrained on VGG-Sound, and SSAST is pretrained on AudioSet and LibriSpeech~\cite{librispeech}.

\noindent \textbf{Audio Processing}. We follow the audio processing of~\cite{kazakos_2021_ICASSP} for extracting the input spectrograms for both models, noting that this outperformed the default processing of SSAST ($200\times128$ spectrograms for 2s of audio, or $400\times128$ for 4s of audio sampled at 16kHz). Namely, audio is resampled at 24kHz for both models. We randomly sample 2s of audio to create log-mel-spectrograms with 128 Mel bands. If the audio annotation is shorter than 2s we pad the produced spectrogram with its last column. We use a window and hop size of 10ms and 5ms respectively, resulting in a spectrogram of size $400 \times 128$.

\noindent\textbf{Training \& Validation Configuration.} We train both models for 30 epochs, setting the initial learning rate to $1\mathrm{e}{-3}$ for ASF which decays to 10\% on epoch 25 and $1\mathrm{e}{-4}$ for SSAST, which is warmed up from $1\mathrm{e}{-6}$ for 2 epochs and decays to 5\% then 1\% on epochs 10 and 20. Both models are trained with cross-entropy loss, optimising ASF using SGD with Nesterov momentum equal to 0.9, and SSAST using AdamW with $(\beta_{1}, \beta_{2})=(0.9, 0.999)$. Both models use a weight decay of $1\mathrm{e}{-4}$ and a batch size of 128. We use a base $384\times384$ ViT with patch size 16 as the backbone for SSAST and the $8\times8$ ResNet50 variant of ASF. For data augmentation, SpecAugment~\cite{Park_2019} is used, again following~\cite{kazakos_2021_ICASSP}, using two frequency masks with $F=27$, two time masks with $T=25$ and time warp with $W=5$. 
We use test augmentations similar to~\cite{kazakos_2021_ICASSP},  dividing the audio into 5 equally sized sub-clips and then averaging their individual predictions from the networks.  For the linear probe results, we freeze the backbones of SSAST and ASF and train only the last linear layer with the same training hyperparameters and pretrained backbones as before.

\begin{figure}[t]
\centering
\begin{subfigure}[t]{0.49\linewidth}
    \centering
    \includegraphics[width=\linewidth]{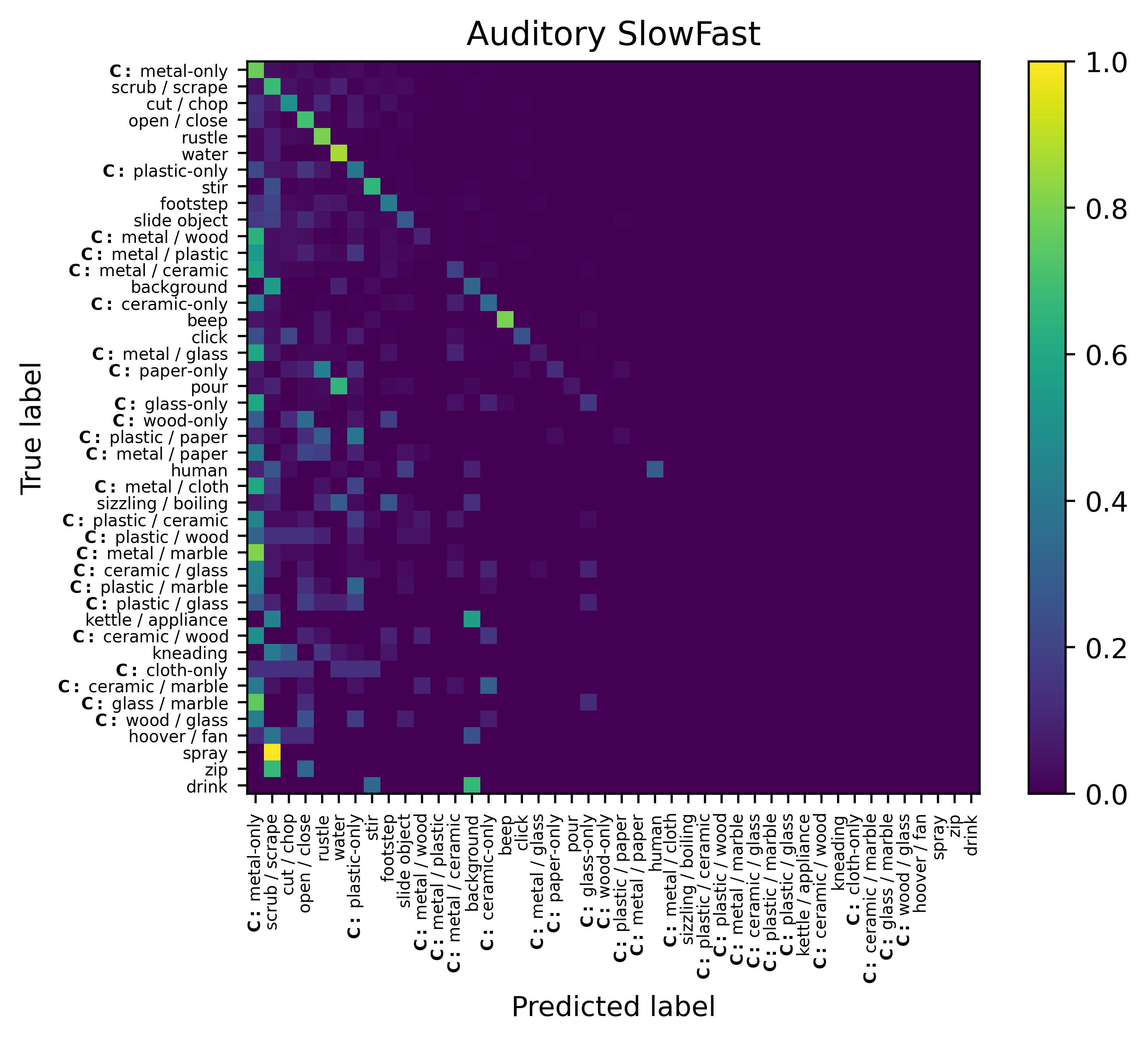}
    \caption{}
   \label{fig:asf_confusion} 
\end{subfigure}
\hfill
\begin{subfigure}[t]{0.49\linewidth}
    \centering
   \includegraphics[width=\linewidth]{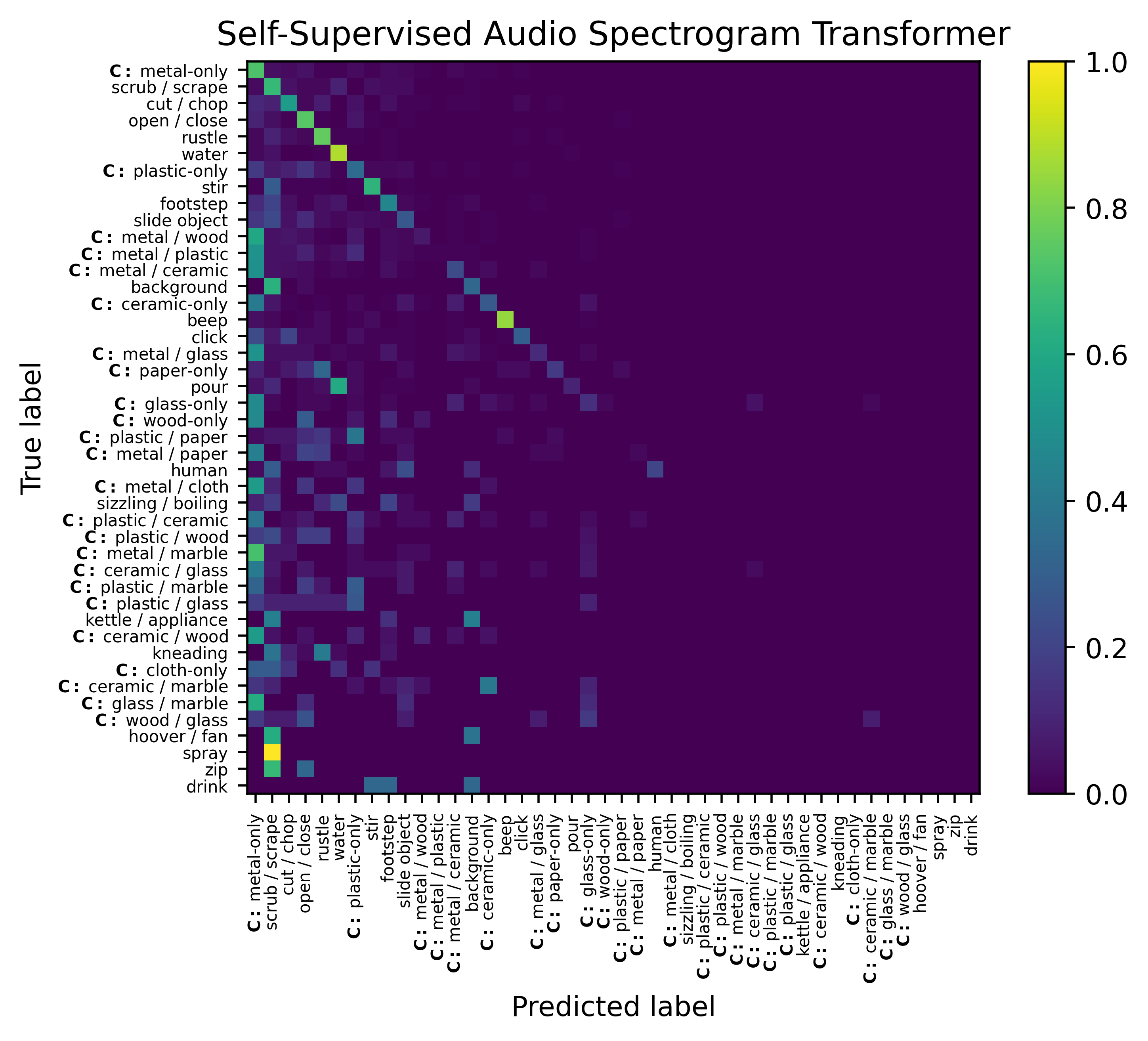}
    \caption{}
   \label{fig:ssast_confusion}
\end{subfigure}
\begin{subfigure}[t]{0.49\linewidth}
    \centering
   \includegraphics[width=\linewidth]{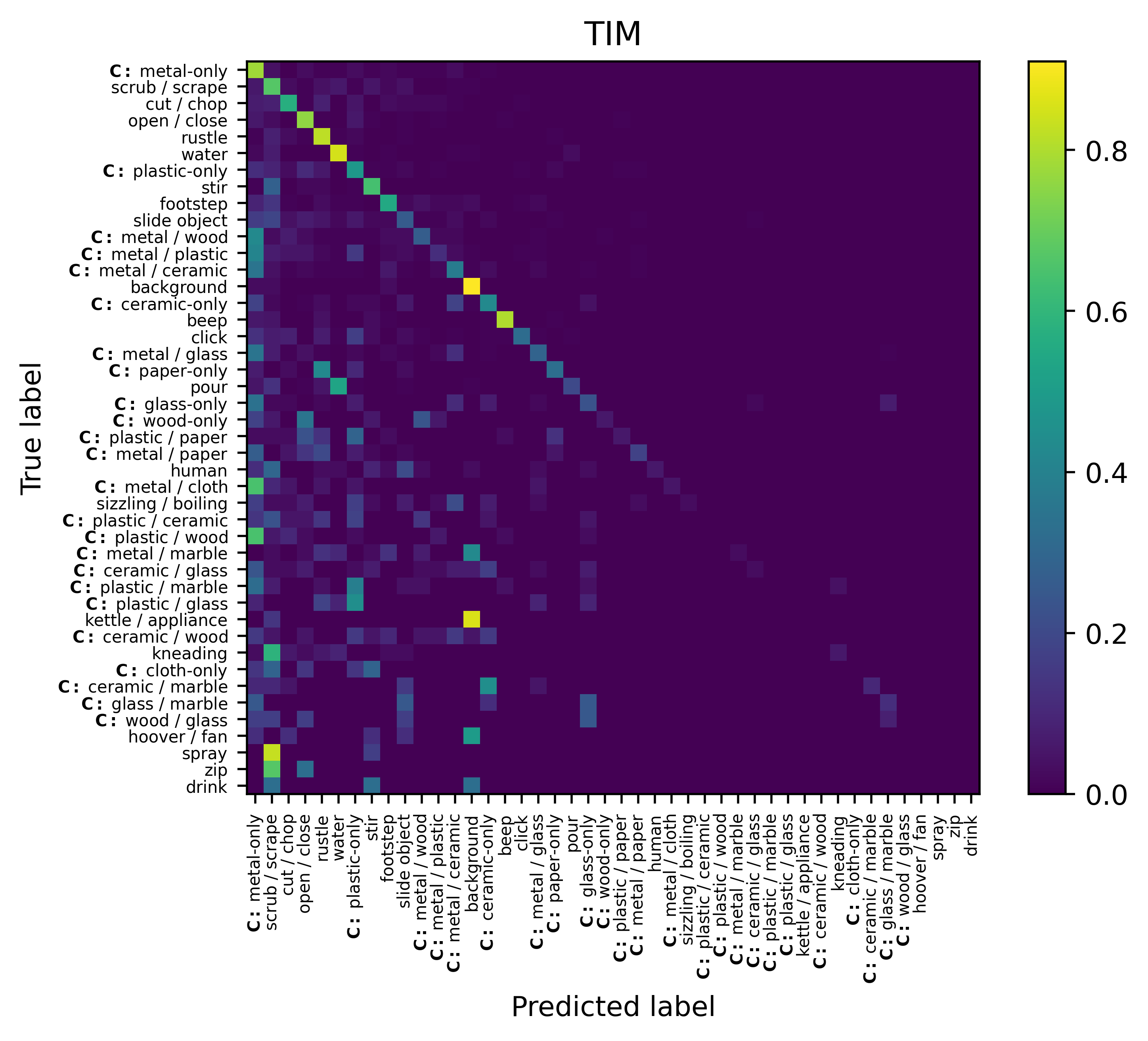}
    \caption{}
   \label{fig:tim_confusion}
\end{subfigure}
\hfill
\begin{subfigure}[t]{0.49\linewidth}
    \centering
    \includegraphics[width=\linewidth]{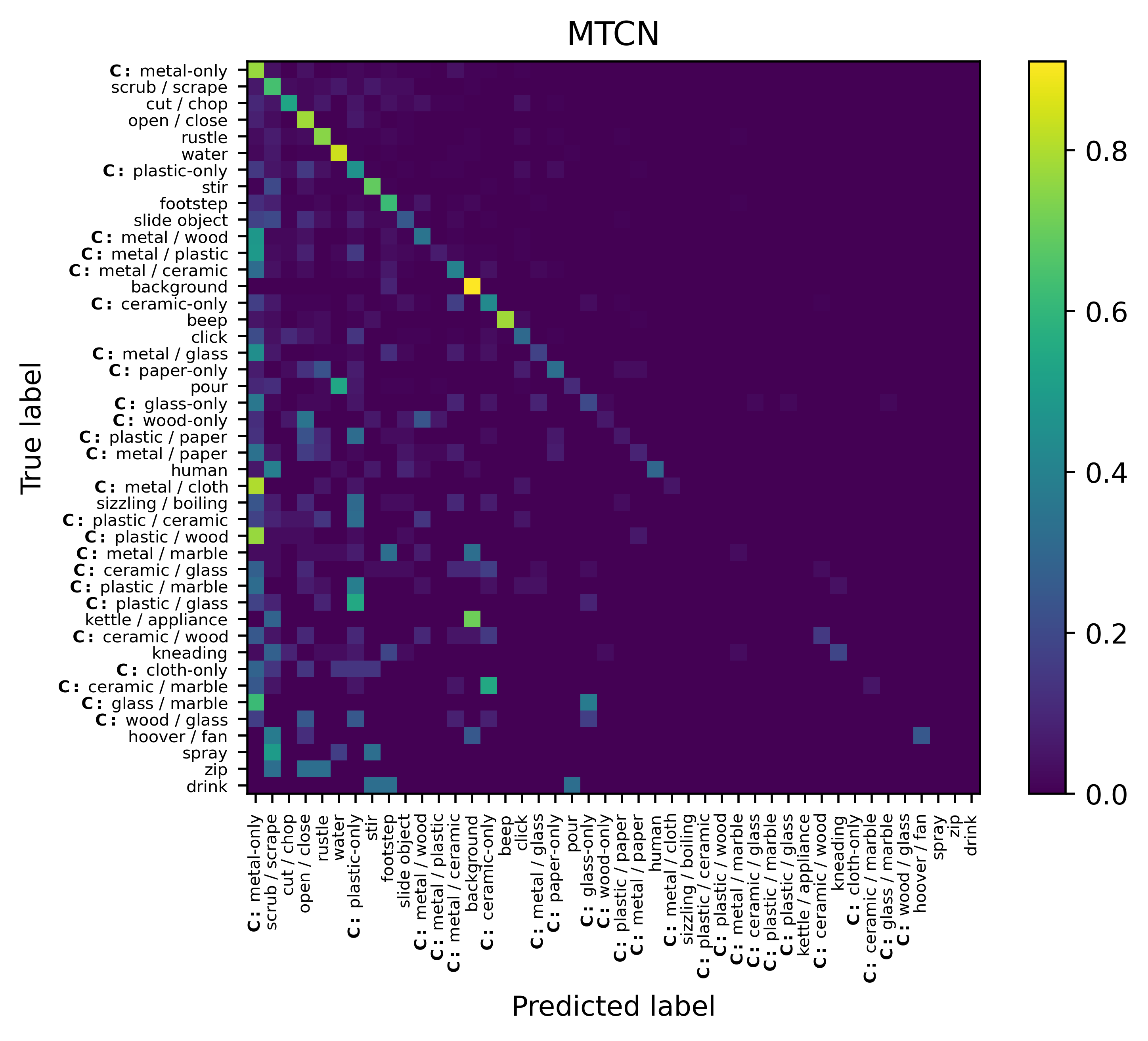}
    \caption{}
   \label{fig:mtcn_confusion} 
\end{subfigure}
\caption{Confusion Matrices on the validation set for: a) ASF, b) SSAST,  c) TIM and d) MTCN.}
\label{fig:confusion}
\end{figure}

\noindent\textbf{Results.} We report quantitative results for both models in Table~\ref{tab:quantresults}. Overall, ASF outperforms SSAST by 0.28\%, 0.74\% and 1.11\% for top-1 accuracy on the validation, recognition test and entire test set respectively. ASF exhibits better mAP for the validation set  and recognition test set,  whereas SSAST performs better on the entire test set, suggesting these models share a similar level of robustness to the long-tailed data. The performance of the linear probe drops significantly compared to fine-tuning results for ASF and almost halves for SSAST. In the latter case, we note that self-supervision alone does not learn class-discriminative features.

Figure~\ref{fig:asf_confusion} and Figure~\ref{fig:ssast_confusion} show the validation confusion matrices for fine-tuned ASF and SSAST respectively. We see that both models are able to detect a subset of distinctive, unique sounds such as rustle, water and beep. Concerning the collision-based classes, both models tend to classify uni-material collisions more successfully than bi-material collisions, but generally produce a false positive prediction of the metal-only collision class, suggesting that the models may struggle to detect how material properties alter the sound produced from a collision.

We also perform a small-scale evaluation of human performance on audio-only sound recognition to compare human perception with computational models. We sample up to 10 clips per class from the \datasetname\ validation set, resulting in 412 clips nearly-balanced subset.
Using the LISA interface (Figure~\ref{fig:lisa_correction_interface}), annotators (i.e. the paper's authors) classify each clip using the \datasetname\ label set without access to video.
Table~\ref{tab:human_performance} shows the results. Humans achieve 20.8\% accuracy, close to ASF (21.1\%) and above SSAST (19.7\%), aligning with the mCA in Table~\ref{tab:quantresults} due to balanced sampling. Notably, humans are especially effective at recognising non-collision sounds, achieving 36.2\% accuracy, which surpasses ASF by 1.7\% and SSAST by 3.5\%.

\begin{table}[t]
\footnotesize
\centering
\caption{Human performance comparing to audio-only recognition baselines on a subset of the \datasetname\ validation set. We report the overall performance as well as performance on both collision and non-collision sounds.}
\begin{tabular}{@{}cccc@{}}
\toprule
               & \textbf{Total} & \textbf{Collision} & \textbf{Non-collision} \\ \midrule
Human & 20.8\%         & 9.4\%              & 36.2\%                 \\
ASF~\cite{kazakos_2021_ICASSP}   &  21.1\%             &  11.6\%                   &          34.5\%              \\
SSAST~\cite{SSAST} &     19.7\%           &      9.8\%              &           32.7\%             \\ \bottomrule
\end{tabular}
\label{tab:human_performance}
\end{table}

\subsection{Audio-Visual Sound Recognition}
Here, we introduce the visual modality and assessing the impact on sound recognition performance. 

\noindent \textbf{Baselines.} For the Audio-Visual baselines, we use MTCN~\cite{kazakos2021MTCN} and TIM~\cite{Chalk_2024_CVPR}.

\noindent \textbf{Audio-Visual Processing.} For TIM, we extract dense, overlapping features using ASF as the backbone. We first fine-tune the backbone for recognition, randomly sampling 1s of audio to create log-mel spectrograms of shape  $200 \times 128$ and then we extract features representing 1 second of audio every 0.2 seconds for each video in \datasetname. For the visual modality, we extract features at the same density as the audio ones, using Omnivore~\cite{girdhar2022omnivore} as the backbone, which has been pre-trained on EPIC-KITCHENS-100. These features then create the transformer input sequences for the model.
For MTCN, we use same Omnivore backbone for visual and ASF backbone for auditory features for a fair comparison with TIM.
We extract 10 features for both audio and visual within each action temporal segments.

\noindent \textbf{Training \& Validation Configuration.} We train TIM in the same way as~\cite{Chalk_2024_CVPR}, though we modify the augmentation strategy by sampling all input features for a given window from the same augmented feature set, as it showed improved performance. 
For MTCN, we follow the same configuration described in the original paper~\cite{kazakos2021MTCN}.

\noindent \textbf{Results.} We report the audio-visual results in Table~\ref{tab:quantresults}. Here, we see that the visual modality assists in audio-based interaction recognition, where both audio-visual baselines consistently improve across metrics. TIM outperforms MTCN on Top-1 accuracy on the validation set and mAP on the entire test set, with MTCN performing better in the remaining metrics. As MTCN is given neighbouring action start-end times during inference, it can exploit the relationships between neighbouring sound actions, especially for repetitive sounds.

We also show the confusion matrices in Figure~\ref{fig:tim_confusion} and Figure~\ref{fig:mtcn_confusion}. When introducing the visual modality, we see a better accuracy for the collision classes, as the models are now able to exploit the visual appearance of colliding objects to better distinguish their sounds. MTCN in particular is also able to better classify tail classes such as hoover / fan, kneading and ceramic / wood collisions, again due to the clear visual signifiers these audible actions produce.

\begin{figure*}[t]
    \centering
    \includegraphics[width=\linewidth]{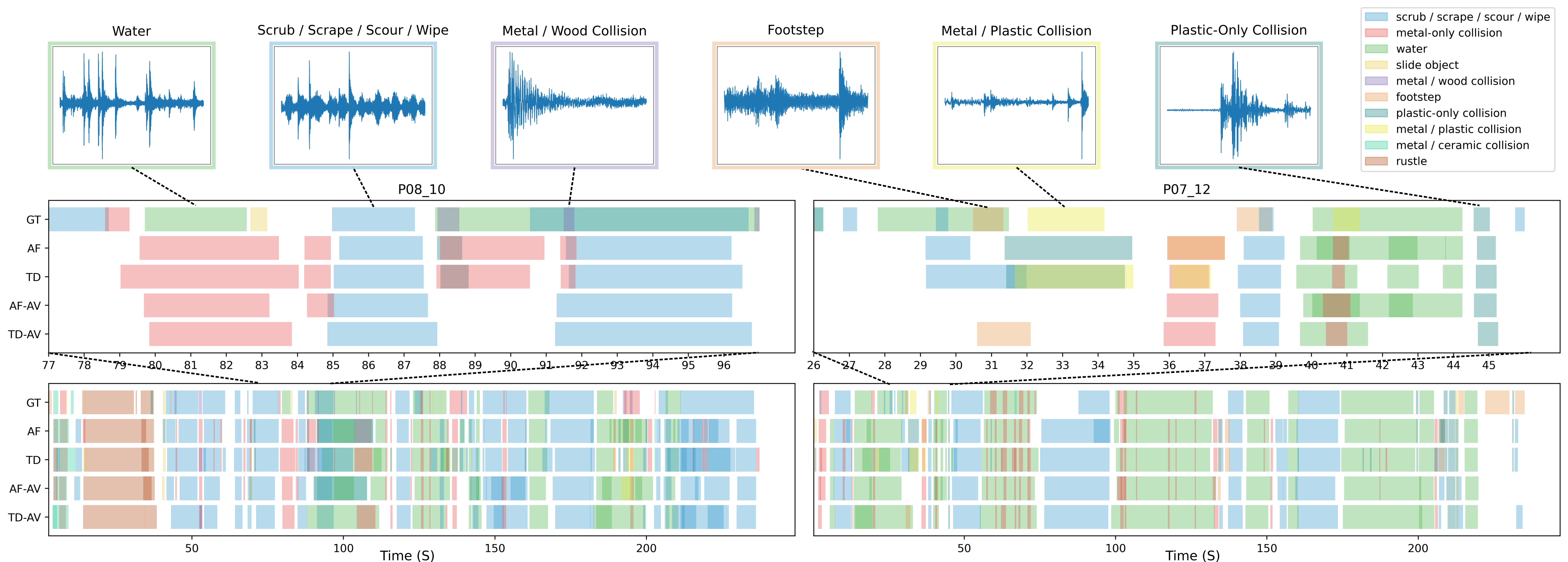}
    \caption{Qualitative sound detection results for \datasetname\ . Here, we show only the top-10 most occurring sound classes across the two selected videos P08\_10 (left) and P07\_12 (right) from the validation set. At the top, we display waveforms for selected audio interactions. Below, we then show the ground-truth (GT), as well as the predictions for audio-only Actionformer~(AF), audio-only TriDet (TD) and their audio-visual counterparts (AF-AV, TD-AV) for the full video (bottom) as well as zoomed-in 20 second region (middle).
    }
    \label{fig:detection_figure}
\end{figure*}

\subsection{Audio-only Sound Detection}
In this section, we train and evaluate audio-only state-of-the-art sound detection models on \datasetname.

\noindent \textbf{Baselines.} We use ActionFormer~\cite{zhang2022actionformer} and TriDet~\cite{shi2023tridet}.

\noindent \textbf{Audio Processing.} For both models, we use the same audio features as used for the TIM baseline.

\noindent \textbf{Training \& Validation Configuration.} We train each model for 16 epochs, using a learning rate of $1\mathrm{e}{-4}$, which is warmed up for 5 epochs before following a cosine annealing decay scheduler. The model is trained using a sigmoid focal loss~\cite{lin2020focalloss} for classification and a centralised distance IoU loss~\cite{zheng2020diou} for regression. The optimiser used is AdamW with $(\beta_{1}, \beta_{2})=(0.9, 0.999)$, a weight decay of $0.05$ and a batch size of 2. 

\noindent \textbf{Results.} We report detection results in Table~\ref{tab:sota_epic_sounds_test_det}. 
In audio-only, TriDet outperforms ActionFormer by 0.1 average mAP on both the detection test set and the entire test set, whereas for the validation set, ActionFormer outperforms TriDet by 0.2 average mAP. 
Typically, ActionFormer appears to be more accurate with its regressed proposals, showing higher average precision at the strictest 0.5 IoU threshold for all splits. 
We show qualitative results for both TriDet and ActionFormer for two videos in Figure~\ref{fig:detection_figure}, along with a zoomed-in crop of a dense 20-second window in the video. 
Here, we show the predictions for the top-10 most frequent audio classes across the two videos, highlighting how the detection baselines can distinguish between multiple, potentially overlapping sounds. 

\begin{table}[t]
    \centering
        \caption{ Results of the Baseline Models on the \datasetname\ validation, detection test and entire test splits. We report the average precision at IoU thresholds $[0.1, 0.2, 0.3, 0.4, 0.5]$ and their average across all thresholds.}
    \resizebox{\linewidth}{!}{
        \begin{tabular}{| c | c | c c c c c | c |}            \hline
             \multirow{2}{*}{\textbf{Split}} & \multirow{2}{*}{\textbf{Method}} & \multicolumn{6}{c|}{\textbf{Average Precision (AP)}} \\
             \cline{3-8} & & @0.1 & @0.2 & @0.3 & @0.4 & @0.5 & Avg. \\
             \hline 
             \multirow{4}{*}{\textbf{Validation}}  & ActionFormer~\cite{zhang2022actionformer} & 16.5 & 15.2 & 13.7 & 12.0 & 10.1 & 13.5 \\
             & TriDet~\cite{shi2023tridet} & 16.1 & 14.9 & 13.6 & 11.9 & 10.0 & 13.3 \\
             & ActionFormer-AV~\cite{zhang2022actionformer} & 18.2 & 17.1 & \textbf{15.1} & 12.1 & 10.0 & 14.5 \\ 
             & TriDet-AV~\cite{shi2023tridet} & \textbf{18.6} & \textbf{17.3} & \textbf{15.1} & \textbf{12.7} & \textbf{10.2} & \textbf{14.8} \\
             \hline
             \multirow{4}{*}{\textbf{Detection Test}} & ActionFormer~\cite{zhang2022actionformer} & 16.4 & 14.6 & 12.6 & 10.6 & 8.5 & 12.5 \\
             & TriDet~\cite{shi2023tridet} & 16.6 & 14.7 & 12.7 & 10.6 & 8.3 & 12.6 \\
             & ActionFormer-AV~\cite{zhang2022actionformer} & \textbf{17.3} & \textbf{15.7} & \textbf{13.7} & \textbf{11.8} & \textbf{9.7} & \textbf{13.6} \\ 
             & TriDet-AV~\cite{shi2023tridet} & 17.1 & 15.3 & 13.2 & 11.0 & 8.6 & 13.0 \\
            \hline
             \multirow{4}{*}{\textbf{Entire Test}} & ActionFormer~\cite{zhang2022actionformer} & 15.2 & 13.4 & 11.6 & 9.6 & 7.6 & 11.5 \\
             & TriDet~\cite{shi2023tridet} & 15.4 & 13.7 & 11.8 & 9.8 & 7.5 & 11.6 \\
             & ActionFormer-AV~\cite{zhang2022actionformer} & \textbf{16.0} & \textbf{14.5} & \textbf{12.6} & \textbf{10.7} & \textbf{8.6} & \textbf{12.5} \\ 
             & TriDet-AV~\cite{shi2023tridet} & 15.8 & 14.2 & 12.3 & 10.2 & 7.9 & 12.1 \\
            \hline
        \end{tabular}
    }
    \label{tab:sota_epic_sounds_test_det}
\end{table}

\subsection{Audio-Visual Sound Detection}
Again, we extend the previous challenge by evaluating on audio-visual models. 

\noindent \textbf{Baselines.} For the baseline, we train visual-counterparts of both ActionFormer and TriDet, combining their predictions with the audio-version to regress the action boundaries and classify the ongoing sound within these boundaries.

\noindent \textbf{Audio-Visual Processing.} This mathces the method described for the TIM recognition baseline. 

\noindent \textbf{Training \& Validation Configuration.} The visual models are trained in the same way as their audio-counterparts, with visual input features. 
To create a single set of proposals, we combine the predictions of each time step from the audio and visual models. 

We follow~\cite{zhang2022actionformer} and re-weight the confidence $\mathbf{p}(\cdot)$ and action boundaries $\mathbf{d}(\cdot)$ of each proposal by:

\begin{equation}
\begin{split}
    \mathbf{p}(interaction)\ &=\ \mathbf{p}(audio)^{\alpha} \ \mathbf{p}(visual)^{(1-\alpha)} \\
    \mathbf{d}(interaction)\ &=\ \omega \mathbf{d}(audio) +  (1-\omega) \mathbf{d}(visual) \\
    \omega &= \mathbf{p}(audio) / (\mathbf{p}(audio)+\mathbf{p}(visual)) \\
\end{split}
\end{equation}
where $\alpha\! = \! 0.8$ for ActionFormer and $\alpha\! = \! 0.7$ for TriDet.
These hyperparameters achieve the best performance.
The high values of $\alpha$ allow the models to be predominantly guided by the audio modality, but assisted by the visual where necessary. 

\noindent \textbf{Results.} We also report audio-visual detection results in Table~\ref{tab:sota_epic_sounds_test_det}. 
In contrast to audio-only, TriDet-AV now outperforms ActionFormer-AV on the validation set by 0.3 average mAP, whereas ActionFormer-AV outperforms by 0.6 and 0.4 average mAP on the detection test and entire test set respectively. 
In comparison to their audio-only version, TriDet-AV sees an additional boost of 1.5, 0.4 and 0.5 average mAP on the validation, detection and entire test set and ActionFormer-AV exhibits an increase of 1.0, 1.1 and 1.0 average mAP for the same splits. 
The additional performance boost of both models when integrating the visual modality highlights the beneficial information shared between them. 
Again, we show qualitative results for the audio-visual extension of the baselines in Figure~\ref{fig:detection_figure}. 
We see that the inclusion of the visual modality can help eliminate false positive predictions as well as improve the regressed boundaries.

\section{Summary and Impact}
\label{sec:conclusion}
In this paper, we present a large-scale dataset,  \datasetname, which consists of 78.4k categorised segments and 39.2k non-categorised segments, totalling 117.6k segments spanning 100 hours of audio, capturing diverse actions that sound in home kitchens. 
Sound categories are annotated based on audio human descriptions. 
We also provide benchmark performance using the state-of-the-art sound recognition and detection networks. 
The audio annotations in this dataset enable a veridical evaluation of audio classification and detection models, and can replace the current evaluations based on visual annotations. We anticipate that multi-modal approaches will benefit from these audio labels.

\noindent\textbf{Impact.} 
Following the introduction of \datasetname\ in 2023, a number of works have built on this dataset.
We summarise these works here.
\cite{Oncescu24} introduces a method for automatically producing audio-centric captions from video-text datasets using an LLM. 
They use the audio labels provided by \datasetname\ as prompts to generate audio descriptions. \cite{chen2024soundingactions} proposes a self-supervised approach to learn how actions sound. 
To achieve this, they use a novel embedding to reinforce correlations between audio, video, and text modalities. Their method is evaluated on the \datasetname\ recognition challenge. \cite{luo2023difffoley} utilises a method inspired by latent diffusion models for video-to-audio synthesis, to generate high-quality, synchronised audio. They fine-tune and evaluate their method on \datasetname\, showcasing its ability to generate accurate audios for samples such as `open drawer' and `plate clinking'. \cite{Piergiovanni_2024_CVPR} propose a multi-modal, autoregressive model jointly modelling audio-visual information and balancing the learning between the two modalities to produce efficient representations. 
They evaluate their model on the \datasetname\ validation set, producing current SOTA recognition results (79.4\%).

\noindent\textbf{Challenges.} 
Both the recognition and the detection challenges are available for submission on Codalab~\cite{codalab_competitions_JMLR}. 
The recognition challenge received 14 submissions in the first year. 
The winning team introduced AudioInceptionNeXt, motivated by the InceptionNext~\cite{yu2024inceptionnext}, which contains parallel multi-scale
depthwise separable convolutional kernels. 
They achieved 55.43\% top-1 accuracy -- an improvement of +1.46\% and +0.63\% over the SSAST and ASF baselines respectively. 
In 2024, 39 submissions were received with the top score improving to 56.57\%; a further improvement of +1.13\%.
This best performing team employed an ensemble of Auditory SlowFast~\cite{kazakos_2021_ICASSP}, SSAST~\cite{SSAST} and AudioInceptionNext.
They achieved a 56.57\% top-1 accuracy -- an improvement of +1.14\% over the previous year's winner.
This challenge can be found at: \href{https://codalab.lisn.upsaclay.fr/competitions/9729}{https://codalab.lisn.upsaclay.fr/competitions/9729}

The audio detection challenge was established in 2024 and had 34 submissions. 
The top-team scored mAP of 14.82, seeing an improvement of +2.28 mAP over the baseline. 
They trained the model based on ActionFormer~\cite{zhang2022actionformer}, but introduced novel hybrid temporal causal blocking to capture long-range relationships.
The model is implemented under the OpenTAD~\cite{2024opentad} framework.
This detection challenge is available from: \href{https://codalab.lisn.upsaclay.fr/competitions/17921}{https://codalab.lisn.upsaclay.fr/competitions/17921}.

\noindent \textbf{Acknowledgements.} This work proposes a new dataset that is publicly available, and builds on publicly available dataset EPIC-KITCHENS.  Research is supported by EPSRC Doctoral Training Program, EPSRC UMPIRE (EP/T004991/1) and the EPSRC Programme Grant VisualAI (EP/T028572/1). 
This project acknowledges the use of the EPSRC funded Tier 2 facility, JADE-II. 
We also thank Rajan from Elancer and his team, for their huge assistance with annotation.

\bibliographystyle{IEEEbib}
\bibliography{shortstrings, references}

\begin{IEEEbiographynophoto}{Jaesung Huh}
Jaesung Huh received a DPhil degree in Visual Geometry Group, University of Oxford, supervised by Professor Andrew Zisserman. Before that, he used to be a research engineer at Naver Corporation. His research focuses on audio-visual learning and video understanding.
\end{IEEEbiographynophoto}

\begin{IEEEbiographynophoto}{Jacob Chalk}
Jacob Chalk is a Postgraduate Researcher of Computer Vision at the University of Bristol. Jacob is currently funded by the EPSRC Doctoral Training Program, where his research is focused on multi-modal learning, specifically audio-visual methods applied to video understanding. 
\end{IEEEbiographynophoto}

\begin{IEEEbiographynophoto}{Evangelos Kazakos}
Evangelos is a postdoctoral researcher at the Czech Institute of Informatics, Robotics, and Cybernetics (CIIRC CTU). His current research interests lie at the intersection of video and language, with a particular focus on grounding visual concepts in text. Evangelos completed his PhD at the University of Bristol, where his thesis focused on Audio-Visual Egocentric Action Recognition. 
\end{IEEEbiographynophoto}

\begin{IEEEbiographynophoto}{Dima Damen}
Dima Damen is a Professor of Computer Vision at the University of Bristol and Senior Research Scientist at Google DeepMind. Dima is currently an EPSRC Fellow (2020-2025), focusing her research interests in the automatic understanding of object interactions, actions and activities using wearable visual (and depth) sensors. She is the lead for the EPIC-KITCHENS dataset, and has contributed to novel research questions including mono-to-3D, video object segmentation, video domain adaptation, skill/expertise determination from video sequences, dual-domain and dual-time learning as well as multi-modal fusion using vision, audio and language. 
\end{IEEEbiographynophoto}

\begin{IEEEbiographynophoto}{Andrew Zisserman}
Andrew Zisserman is a Royal Society Research Professor and the Professor of Computer Vision Engineering at the Department of Engineering Science at the University of Oxford. His research interests have included multiple view geometry, audio and visual recognition, and large scale retrieval of images and video. His papers have won many best paper and test-of-time awards at international conferences.
\end{IEEEbiographynophoto}

\end{document}